\documentclass[printer]{aa}
\usepackage{graphicx}
\usepackage{txfonts}
\def\lesssim{\mathrel{\hbox{\rlap{\hbox{\lower4pt\hbox{$\sim$}}}\hbox{$<$}}}}
\def\gtrsim{\mathrel{\hbox{\rlap{\hbox{\lower4pt\hbox{$\sim$}}}\hbox{$>$}}}}
\newcommand{\mincir}{\raise -2.truept\hbox{\rlap{\hbox{$\sim$}}\raise5.truept\hbox{$<$}\ }}
\newcommand{\magcir}{\raise -2.truept\hbox{\rlap{\hbox{$\sim$}}\raise5.truept\hbox{$>$}\ }}
\newcommand{\siml}{\raise -2.truept\hbox{\rlap{\hbox{$\sim$}}\raise5.truept\hbox{$<$}\ }}
\newcommand{\simg}{\raise -2.truept\hbox{\rlap{\hbox{$\sim$}}\raise5.truept\hbox{$>$}\ }}
\newcommand{\be}{\begin{equation}}
\newcommand{\ee}{\end{equation}}
\newcommand{\ba}{\begin{eqnarray}}
\newcommand{\ea}{\end{eqnarray}}
\newcommand {\kpc} {$h_{70}^{-1}$ kpc$\;$}

\newcommand {\h} {$h_{70}^{-1}$ Mpc$\;$}
\newcommand {\hh} {$h_{70}^{-1}$ Mpc}
\newcommand {\hhh} {\;h_{70}^{-1} \mathrm{Mpc}}
\newcommand {\ks} {km~s$^{-1} \;$}
\newcommand {\kss} {km~s$^{-1}$}

\newcommand {\mtre} {$\times 10^{13}\;h_{70}^{-1}\;M_{\odot \;}$}
\newcommand {\mtree} {$\times 10^{13}\;h_{70}^{-1}\;M_{\odot}$}
\newcommand {\mqua} {$\times 10^{14}\;h_{70}^{-1}\;M_{\odot} \;$}
\newcommand {\mquaa} {$\times 10^{14}\;h_{70}^{-1}\;M_{\odot}$}
\newcommand {\mqui} {$\times 10^{15}\;h_{70}^{-1}\;M_{\odot} \;$}
\newcommand {\mquii} {$\times 10^{15}\;h_{70}^{-1}\;M_{\odot}$}

\newcommand{\degree}{\ensuremath{\mathrm{^\circ}}}
\newcommand{\arcm}{\ensuremath{\mathrm{^\prime}\;}}
\newcommand{\arcs}{\ensuremath{\arcmm\hskip -0.1em\arcmm \;}}
\newcommand{\arcmm}{\ensuremath{\mathrm{^\prime}}}
\newcommand{\arcss}{\ensuremath{\arcmm\hskip -0.1em\arcmm}}
\newcommand{\dotarcs}{\,\rlap{\hbox{$\mathrm{^\prime\hskip-0.1em^\prime}$}}{\hbox{$.$}}\,}

\newcommand{\dotsec}{\,\rlap{\hbox{$\mathrm{^s}$}}{\hbox{$.$}}\,}
\begin{document}
   \title{The dynamical status of the galaxy cluster Abell 115}
\author{R. Barrena\inst{1}
\and W. Boschin\inst{2,3} 
\and M. Girardi\inst{3,4}
\and M. Spolaor\inst{3,5}
}

   \offprints{R. Barrena, \email{rbarrena@iac.es}}

   \institute{Instituto de Astrof\'{\i}sica de Canarias, C/V\'{\i}a
   L\'actea s/n, E-38200 La Laguna, Tenerife, Spain\\
         \and
              Fundaci\'on Galileo Galilei - INAF, C/Alvarez de Abreu 70,
	      E-38700 Santa Cruz de La Palma, Canary Islands, Spain\\
         \and
              Dipartimento di Astronomia of the Universit\`a degli Studi di 
	      Trieste, via Tiepolo 11, I-34131 Trieste, Italy\\
         \and
              INAF - Osservatorio Astronomico di Trieste, via Tiepolo 11, 
	      I-34131 Trieste, Italy\\
         \and
              Centre for Astrophysics \& Supercomputing, Swinburne University,
              Hawthorn, VIC 3122, Australia\\
              }

\date{Received / Accepted } 

\abstract{} {We present the results of a new spectroscopic and
photometric survey of the hot, binary X--ray cluster A115 at
$z=0.193$, containing a radio relic.}  {Our analysis is based on new
spectroscopic data obtained at the Telescopio Nazionale Galileo for
115 galaxies and on new photometric data obtained at the Isaac Newton
Telescope in a large field. We combine galaxy velocity and position
information to select 85 galaxies recognized as cluster members,
determine global dynamical properties and detect substructures.}  {We
find that A115 appears as a well isolated peak in the redshift space,
with a global line--of--sight (LOS) velocity dispersion $\sigma_{\rm
v}=1362_{-108}^{+126}$ km s$^{-1}$.  Our analysis confirms the
presence of two structures of cluster--type well recognizable in the
plane of the sky and shows that they differ of $\sim 2000$ \ks in the
LOS velocity.  The northern, high velocity subcluster (A115N) is
likely centred on the second brightest cluster galaxy (BCM--A,
coincident with radio source 3C28) and the northern X--ray peak.  The
southern, low velocity subcluster (A115S) is likely centred on the
first brightest cluster galaxy (BCM--B) and the southern X--ray peak.
We estimate that A115S is slightly dynamically more important than
A115N having $\sigma_{\rm v}=900-1100$ \ks vs.  $\sigma_{\rm
v}=750-850$ \kss.  Moreover, we find evidence for two small groups at
low velocities.  We estimate a global cluster virial mass of 2.2--3.5
\mquii.}  {Our results agree with a pre--merging scenario where A115N
and A115S are colliding with a LOS impact velocity $\Delta {{\rm
v}_{\rm rf}} \sim 1600$ \kss.  The most likely solution to the
two--body problem suggests that the merging axis lies at $\sim 20$
degrees from the plane of the sky and that the cores will cross after
$\sim 0.1$ Gyr. The radio relic with its largest dimension
perpendicular to the merging axis is likely connected to this merger.}


\authorrunning{Barrena et al.}
\titlerunning{The dynamical status of A115} 
\maketitle

%

\section{Introduction}
\label{sec:int}

The evolution of clusters of galaxies as seen in numerical simulations
is characterized by the asymmetric accretion of mass clumps from
surrounding filaments (e.g. Diaferio et al. \cite{dia01}). Nearby
clusters are characterized by a variety of morphologies, indicative of
different dynamical properties, elongated distribution, traced by
several galaxy clumps (e.g., Barrena et al. \cite{bar02}; Boschin et
al. \cite{bos06}; Sauvageot et al. \cite{sau05}; Yuan et
al. \cite{yua05}). The presence of substructure is indicative of a
cluster in an early phase of the process of dynamical relaxation or of
secondary infall of clumps into already virialized clusters (see
Girardi \& Biviano \cite{gir02} for a review).

An interesting aspect of these investigations is the possible
connection of cluster mergers with the presence of extended, diffuse
radio sources. These sources are large (up to $\sim1$ \hh), amorphous
cluster sources of uncertain origin and generally steep radio spectra
(Hanisch \cite{han82}; see also and Giovannini \& Feretti \cite{gio02}
for a review). They are classified as halos, if they are
located in the cluster centre, or relics, if they appear in the
peripheral regions of the cluster.  Halos and relics are rare sources
that appear to be associated with very rich clusters that have
undergone recent mergers and thus it has been suggested by various
authors that cluster halos/relics are related to recent merger
activity (e.g., Tribble \cite{tri93}; Burns et al. \cite{bur94};
Feretti \cite{fer99}).

The synchrotron radio emission of halos and relics demonstrates the
existence of large scale cluster magnetic fields, of the order of
0.1--1 $\mu$G, and of widespread relativistic particles of energy
density 10$^{-14}$ -- 10$^{-13}$ erg cm$^{-3}$.  The difficulty in
explaining halos/relics arises from the combination of their large
size and the short synchrotron lifetime of relativistic electrons.
The expected diffusion velocity of the electron population is on the
order of the Alfven speed ($\sim100$ \kss) making it difficult for the
electrons to diffuse over a megaparsec--scale region within their
radiative lifetime. Therefore, one needs a mechanism by which the
relativistic electron population can be transported over large
distances in a short time, or a mechanism by which the local electron
population is reaccelerated and the local magnetic fields are
amplified over an extended region.  The cluster--cluster merger can
potentially supply both mechanisms (e.g., Giovannini et
al. \cite{gio93}; Burns et al. \cite{bur94}; R\"ottgering et
al. \cite{rot94}; see also Feretti et al. \cite{fer02a}; Sarazin
\cite{sar02} for reviews). However, the question is still debated
since the diffuse radio sources are quite uncommon and only recently
we can study these phenomena on the basis of a sufficient statistics
(a few dozens of clusters up to $z\sim0.3$, e.g., Giovannini et
al. \cite{gio99}; see also Giovannini \& Feretti \cite{gio02}; Feretti
\cite{fer05}).

Growing evidence of the connection between diffuse radio emission and
cluster merging is based on X--ray data (e.g., B\"ohringer \&
Schuecker \cite{boh02}; Buote \cite{buo02}). Studies based on a large
number of clusters have found a significant relation between the radio
and the X--ray surface brightness (Govoni et al. \cite{gov01a},
\cite{gov01b}) and between the presence of radio--halos/relics and
irregular and bimodal X--ray surface brightness distribution
(Schuecker et al. \cite{sch01}).  New unprecedented insights into
merging processes in radio clusters are offered by Chandra and
XMM--Newton observations (e.g., Markevitch \& Vikhlinin \cite{mar01};
Fujita et al. \cite{fuj04}).

Optical data are a powerful way to investigate the presence and the
dynamics of cluster mergers (e.g., Girardi \& Biviano \cite{gir02}),
too.  The spatial and kinematical analysis of member galaxies allow us
to detect and measure the amount of substructure, to identify and
analyse possible pre--merging clumps or merger remnants.  This optical
information is really complementary to X--ray information since
galaxies and ICM react on different time scales during a merger (see
numerical simulations by Roettiger et al. \cite{roe97}).
Unfortunately, to date optical data is lacking or poorly exploited and
sparse literature concerns some few individual clusters.  In this
context we have carried on an intensive program to study the
clusters containing extended, diffuse radio emission (Boschin et
al. \cite{bos04}; Boschin et al. \cite{bos06}; Girardi et
al. \cite{gir06}; Barrena et al. \cite{bar07}). In particular, we have
conducted a study of Abell 115 (hereafter A115) having an extended
arc--shape relic (Feretti et al.  \cite{fer84}; Giovannini et
al. \cite{gio87}; Govoni et al. \cite{gov01b}).

A115 is a rich cluster (Abell richness class =3; Abell et
al. \cite{abe89}) known in the literature to be characterized by a
double X--ray peak (A115N and A115S; Forman et al. \cite{for81}) and
by the presence of a strong cooling flow in its northern component
(White et al. \cite{whi97}).  White et al. (\cite{whi97}) give the
following values for the X--ray luminosity and temperature:
$L_\mathrm{X}=14.7$ $\times 10^{44}$, 9.1 $\times 10^{44}$ erg
s$^{-1}$ (in their cosmology) and $T_\mathrm{X}=6.6$, 5.7 keV for the
northern and southern components, respectively. Slightly smaller
temperatures are also found in the literature:
$T_\mathrm{X}=4.9_{-0.6}^{+0.7}$ and 5.2 keV for the northern
component, and $T_\mathrm{X}=5.2_{-1.0}^{+1.4}$ and 4.8 keV for the
southern component (from Shibata et al. \cite{shi99} and Gutierrez \&
Krawczynski \cite{gut05}, respectively).  Both ASCA and Chandra data
show that the temperature of the ICM is highly nonuniform across the
cluster suggesting that the merger is well underway, but not
disturbing the cool subcluster cores (Shibata et al. \cite{shi99};
Gutierrez \& Krawczynski \cite{gut05}). From the optical point of
view, Beers et al. (\cite{bee83}) mapped the galaxy distribution
finding three major clumps of galaxies (A, B, and C in Fig. 4a of
their paper). Northern and southern clumps A and B correspond to the
peaks of the X--ray surface brightness distribution.  However, no
X--ray emission has been found associated with the third eastern
clump.

The northern subcluster contains the very strong radio galaxy 3C28
(Riley \& Pooley \cite{ril75}; Feretti et al. \cite{fer84}; Giovannini
et al. \cite{gio87}). Its host galaxy is the brightest galaxy of the
northern subcluster and is classified as ``elliptical'' by Schombert
\cite{sch87}. The strong X--ray and radio emissions of this cluster
member seem to be explained by the presence of a large amount of hot
gas which is likely to be accreting onto the galaxy. 
This scenario agrees with the observation in the
spectrum of this galaxy of the H$\alpha$--line and other emission
lines, likely tracing a cooling flow system (Crawford et
al. \cite{cra99}).

The diffuse radio source belongs to A115N, and extends from this
sub--cluster to the periphery. According to its non--central location
and its elongated structure, it is classified as a cluster
relic. However, elongated relics are generally at the cluster periphery,
and with the largest dimension roughly perpendicular to the cluster
radial orientation. So this source is quite unusual although several
considerations confirm that it should be a cluster relic (Govoni et
al. \cite{gov01b}).

To date few spectroscopic data are reported in the literature.  Beers
et al. (\cite{bee83}) measured the redshift for 19 cluster members
(see also Zabludoff et al. \cite{zab90} for a later re--reduction with
a new template). In December 2003 we obtained spectra of 115 galaxies
in the cluster region, with the purpose of constraining its dynamical
status.

The plan of this paper is the following. In Sect.~\ref{sec:dat} we
describe our new photometric and spectroscopic observations. In
Sect.~\ref{sec:ana} we provide our results about cluster membership,
global properties, and substructure.  In Sect.~\ref{sec:mass} we draw
our conclusion about cluster structure and estimate cluster mass.  In
Sect.~\ref{sec:merg} we discuss our results and posit a plausible
scenario for the dynamical status of A115. We summarize our results in
Sect.~\ref{sec:sum}.

Throughout this paper, we use $H_0=70\,\,h_{70}$ km s$^{-1}$ Mpc$^{-1}$ in a
flat cosmology with $\Omega_{\rm m}=0.3$ and $\Omega_{\Lambda}=0.7$. In the
adopted cosmology, 1\arcm corresponds to $\sim 192$ \kpc at the
cluster redshift.

\section{Data}
\label{sec:dat}

\subsection{Spectroscopy}
\label{sec:spe}

Multi--object spectroscopic observations of A115 were carried out at
the TNG telescope in December 2003 during the program of proposal
AOT8/CAT--G6. We used DOLORES/MOS with the LR--B Grism 1, yielding a
dispersion of 187 \AA/mm, and the Loral CCD of $2048\times2048$ pixels
(pixel size of 15 $\mu$m). This combination of grating and detector
results in dispersions of 2.8$\,$\AA/pix. We have taken five MOS masks
for a total of 152 slits. We acquired three exposures of 1800 s for
each masks. Wavelength calibration was performed using Helium--Argon
lamps. Reduction of spectroscopic data was carried out with IRAF
\footnote{IRAF is distributed by the National Optical Astronomy
Observatories, which are operated by the Association of Universities
for Research in Astronomy, Inc., under cooperative agreement with the
National Science Foundation.} package.

Radial velocities were determined using the cross--correlation
technique (Tonry \& Davis \cite{ton79}) implemented in the RVSAO
package (developed at the Smithsonian Astrophysical Observatory
Telescope Data Center).  Each spectrum was correlated against six
templates for a variety of galaxy spectral types: E, S0, Sa, Sb, Sc,
Ir (Kennicutt \cite{ken92}).  The template producing the highest value
of $\cal R$, i.e., the parameter given by RVSAO and related to the
signal--to--noise of the correlation peak, was chosen.  Moreover, all
spectra and their best correlation functions were examined visually to
verify the redshift determination.  In some cases (IDs 85, 112 and  115,
see Table \ref{tab1}) we took the EMSAO redshift as a reliable
estimate of the redshift.

For eleven galaxies we obtained two redshift determinations, which are
of similar quality.  This allow us to obtain a more rigorous estimate
for the redshift errors since the nominal errors as given by the
cross--correlation are known to be smaller than the true errors (e.g.,
Malumuth et al.  \cite{mal92}; Bardelli et al. \cite{bar94}; Ellingson
\& Yee \cite{ell94}).  We fit the first determination vs. the second
one by using a straight line and considering errors in both
coordinates (e.g., Press et al. \cite{pre92}). The fitted line agrees
with the one to one relation, but, when using the nominal
cross--correlation errors, the small value of $\chi^2$--probability
indicates a poor fit, suggesting the errors are underestimated.  Only
when nominal errors are multiplied by a $\sim 1.3$ factor the observed
scatter can be explained. Therefore, hereafter we assume that true
errors are larger than nominal cross--correlation errors by a factor
1.3. For the eleven galaxies we used the average of the two redshift
determinations and the corresponding error.

Our spectroscopic survey consists of 115 galaxies taken in a
field of $\sim$ 15\arcm $\times$ 20\arcm.  

We also determined the equivalent widths (EW hereafter) of the
absorption line H$\delta$ and the emission line [OII], in order to
classify post--starburst and starburst galaxies. We estimated the
minimum measurable EW of each spectrum as the width of a line spanning
2.8 \AA ~(our dispersion) in wavelength, with an intensity three times
the rms noise in the adjacent continuum.  This estimation yields upper
measurable limits of $\sim 4.5$ \AA ~in EW.

We use a conservative approach leading to a sparse spectral
classification ($\sim 50$\% of the sample, see Table~\ref{tab1}).  We
follow the classification by Dressler et al. (\cite{dre99}; see also
Poggianti et al. \cite{pog99}).  We define ``e''--type galaxies those
showing active star formation as indicated by the presence of an [OII]
and, in particular, ``e(b)'' galaxies when the equivalent width of
EW([OII]) is $\le-40$ \AA $\ $ (likely starburst galaxies);
``e(a)''--type galaxies those having EW(H$\delta$)$\ge$4 \AA$\,$;
"e(c)"--type galaxies those having moderate emission lines and
EW(H$\delta$)$<$4 \AA$\,$ (likely spiral galaxies).  We define ``k+a''
and ``a+k''--type galaxies those having $4\le$EW(H$\delta$)$\le$8 \AA
$\,$ and EW(H$\delta$)$>$8 \AA, respectively, and no emission lines
(the so called ``post--starbust''). Moreover, out of galaxies having
the cross--correlation coefficient ${\cal R}\gtrsim 7$ --
corresponding to $S/N \gtrsim 10$ as obtained using A773 data (Barrena
et al. \cite{bar07}) -- we define ``k''--type galaxies those having
EW(H$\delta$)$<$3 \AA $\,$ and no emission lines (likely passive
galaxies). We classify 62 galaxies finding 38 ``passive'' galaxies, 24
``active'' galaxies (i.e., 14 ``k+a''/``a+k'' and 10 ``e'' galaxies).

\subsection{Photometry}
\label{sec:pho}
As far as photometry is concerned, our observations were carried out
with the Wide Field Camera (WFC), mounted at the prime focus of the
2.5m INT telescope (located at Roque de los Muchachos observatory, La
Palma, Spain). We observed A115 in 18 December 2004 in photometric
conditions with a seeing of about 2\arcss.

The WFC consists of a 4 chip mosaic covering a 30\arcmm$\times$30\arcm
field of view, with only a 20\% marginally vignetted area. We took 10
exposures of 720 s in $B_{\rm H}$ and 360 s in $R_{\rm H}$ Harris
filters (a total of 7200 s and 3600 s in each band) developing a
dithering pattern of ten positions. This observing mode allowed us to
build a master ``supersky'' image that was used to correct our images
for vignetting and fringing patterns (Gullixson \cite{gul92}). In
addition, the dithering helped us to clean cosmic rays and avoid gaps
between CCD chips. The complete reduction process (including flat
fielding, bias subtraction and bad--columns elimination) yielded a
final co-added image where the variation of the sky was lower than
0.4\% in the whole frame. Another effect associated with the wide
field frames is the distortion of the field. In order to match the
photometry of several filters (in our case, only $B_{\rm H}$ and
$R_{\rm H}$), a good astrometric solution taking into account these
distortions is needed. Using IRAF tasks and taking as reference the
USNO B1.0 catalog we were able to find an accurate astrometric
solution (rms$\sim0.5''$) across the full frame. The photometric
calibration was performed using Landolt standard fields achieved
during the observation.

We finally identified galaxies in our $B_{\rm H}$ and $R_{\rm H}$
images and measured their magnitudes with the SExtractor package
(Bertin \& Arnouts \cite{ber96}) and AUTOMAG procedure. In few cases,
(e.g., close companion galaxies, galaxies close to defects of CCD),
the standard SExtractor photometric procedure failed. In these cases
we computed magnitudes by hand. This method consists in assuming a
galaxy profile of a typical elliptical and scale it to the maximum
observed value. The integration of this profile give us an estimate of
the magnitude. The idea of this method is similar to the PSF
photometry, but assuming a galaxy profile, more appropriate in this
case.

We transformed all magnitudes into the Johnson--Cousins system (Johnson
\& Morgan \cite{joh53}; Cousins \cite{cou76}). We used $B=B\rm_H+0.13$
and $R= R\rm_H$, as derived from the Harris filter characterization
(http://www.ast.cam.ac.uk/$\sim$wfcsur/technical/photom/colours/) and
assuming a $B$--$V\sim1.0$ for E--type galaxies (Poggianti
\cite{pog97}). As a final step, we estimated and corrected the
galactic extinction $A_B \sim 0.25$, $A_R \sim 0.15$ from Burstein \&
Heiles (\cite{bur82}) reddening maps. We estimated that our
photometric sample is complete down to $B=19.5$ (21.0) and $R=22.0$
(23.0) for $S/N=5$ (3) within the observed field.

We assigned $B$ and $R$ magnitudes to the whole spectroscopic sample.
We measured redshift for galaxies down to $R\sim$ 20.5 mag, but we are
complete to 60\% down to $R=$18 mag [within a region of 13\arcm
$\times$ 20 \arcm around
R.A.=$00^{\mathrm{h}}56^{\mathrm{m}}02\dotsec0$,
Dec.=$+26\degree\,23\arcm\,00\arcs$ (J2000.0)].

Table~\ref{tab1} lists the velocity catalogue (see also
Fig.~\ref{figimage} and Fig.~\ref{figisofote}): identification number
of each galaxy, ID (Col.~1); ID code following the IAU nomenclature
(Col.~2); right ascension and declination, $\alpha$ and $\delta$
(J2000, Col.~3); $B$ and $R$ magnitudes (Cols.~4 and 5, respectively);
heliocentric radial velocities, ${\rm v}=cz_{\sun}$ with errors,
$\Delta {\rm v}$ (Cols.~6 and 7, respectively); spectral
classification SC (Col.~8).

\section{Analysis \& Results}
\label{sec:ana}

\subsection{Member selection and global properties}
\label{sec:mem}

To select cluster members out of the 115 galaxies having redshifts, we
follow the two steps procedure.  First, we perform the
adaptive--kernel method (hereafter DEDICA, Pisani \cite{pis93} and
\cite{pis96}; see also Fadda et al. \cite{fad96}; Girardi et
al. \cite{gir96}; Girardi \& Mezzetti \cite{gir01}). We find the
significant peaks in the velocity distribution $>$99\% c.l..  This
procedure detects A115 as a one--peak structure at $z\sim0.1937$
populated by 88 galaxies considered as candidate cluster members (see
Fig.~\ref{fighisto}).  Out of non--member galaxies, 12 and 15 are
foreground and background galaxies, respectively.

\addtocounter{figure}{+2}
\begin{figure}
\centering
\resizebox{\hsize}{!}{\includegraphics{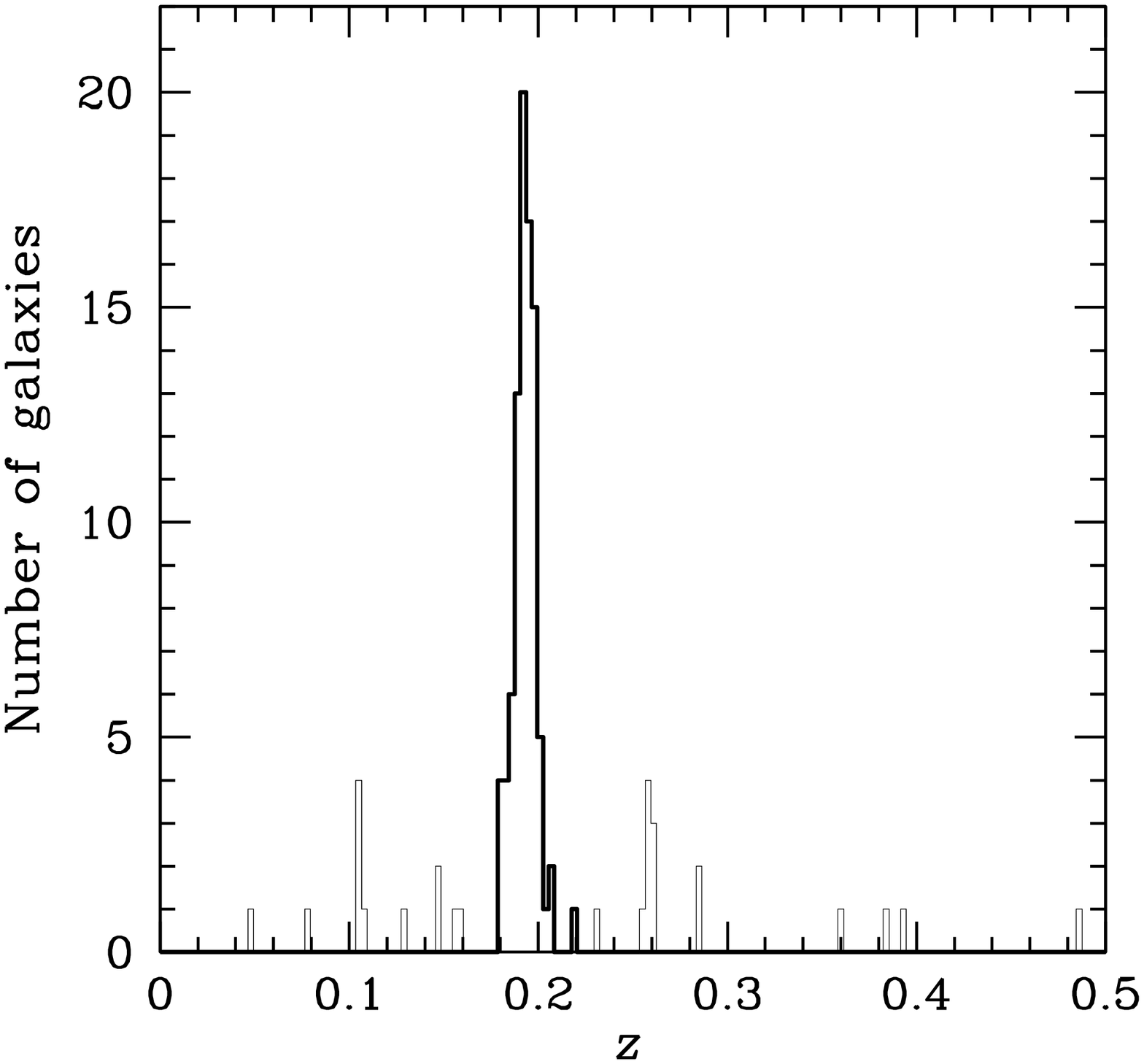}}
\caption
{Redshift galaxy distribution. The solid line histogram refers to
galaxies assigned to the cluster according to the DEDICA
reconstruction method.}
\label{fighisto}
\end{figure}

\begin{figure}
\centering 
\resizebox{\hsize}{!}{\includegraphics{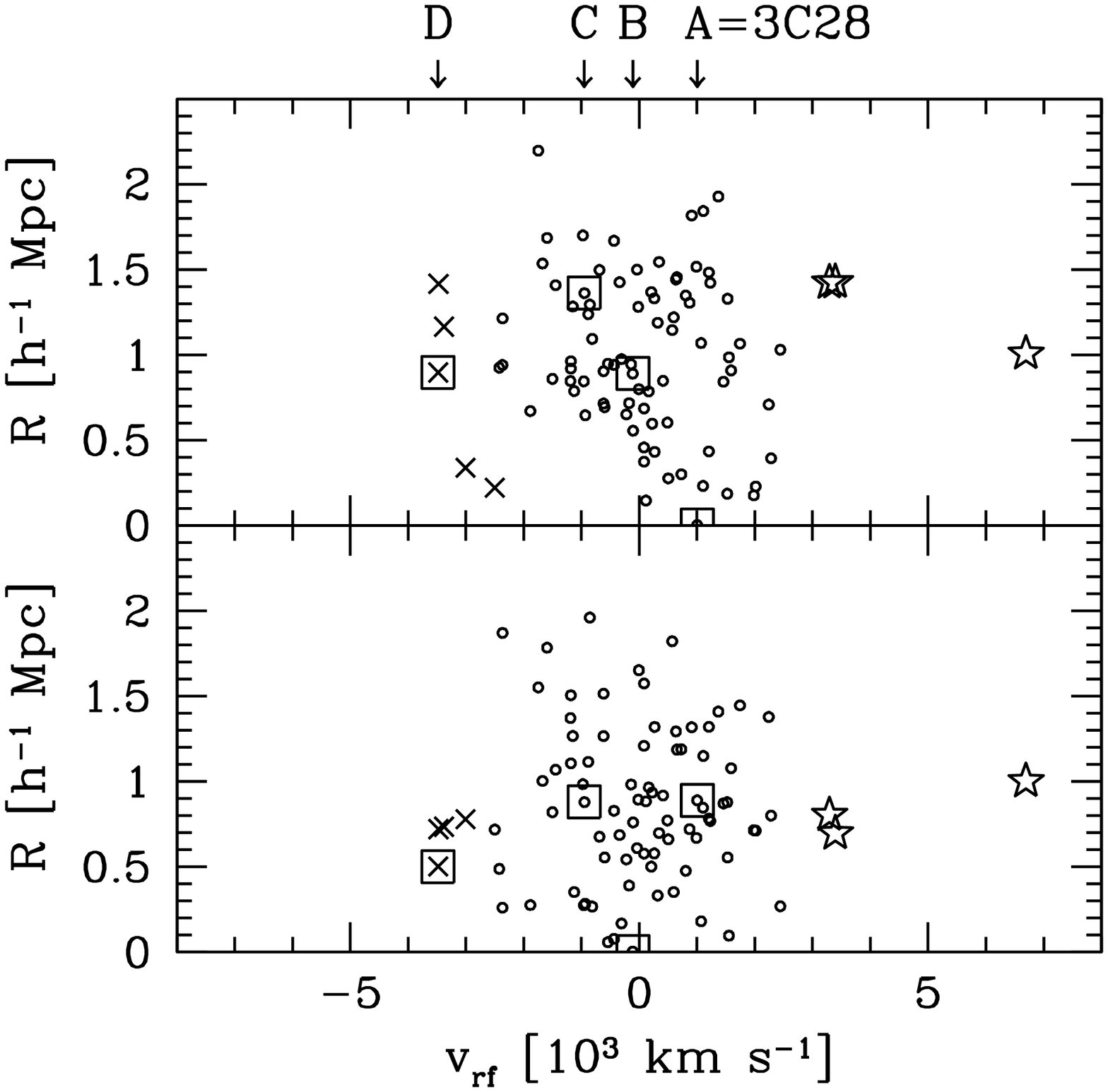}}
\caption{Projected clustercentric distance vs. rest--frame velocity
v$_{\rm rf}$=(v-$<$v$>$)$/(1+z)$ for the 88
galaxies in the main peak (see Fig.~\ref{fighisto}) showing galaxies
detected as interlopers by our ``shifting gapper'' procedure (stars
and crosses) when using BCM--A and --B as alternative cluster centres
({\em top and bottom panels}, respectively). We reject the three
galaxies with highest velocities (stars) to build our basic sample of
85 galaxies and all galaxies indicated by crosses and stars in the top
panel to build our Sample2.  The three brightest cluster members each
one corresponding to Beers et al. (\cite{bee83}) clumps (BCM--A, --B
and --C) and a fourth bright galaxy (BCM--D) are indicated by
squares.}
\label{figvd}
\end{figure}

The projected clustercentric distance vs. the rest-frame velocity of
the 88 galaxies is shown in Fig.~\ref{figvd} where we also show
the three brightest galaxies in our sample, each of
them corresponding to a different galaxy density clumps as found by
Beers et al. (\cite{bee83}), i.e. IDs~54, 21(=3C28) and 104 (hereafter
BCM--B, BCM--A, and BCM--C) corresponding to B (A115S), A (A115N), and
C clumps, respectively.  The position of BCM--B is very close to that
of the southern X--ray peak
[R.A.=$00^{\mathrm{h}}55^{\mathrm{m}}58\dotsec81$,
Dec.=$+26\degree\,19\arcm\,58\dotarcs9$ (J2000.0) as recovered from
X--ray Chandra data, see Fig.~\ref{figisofote}].  BCM--A=3C28 is
coincident with the northern X--ray peak and is very close to the
position of the X--ray centroid when masking the X--ray emission of
the point source [R.A.=$00^{\mathrm{h}}55^{\mathrm{m}}53\dotsec2$,
Dec.=$+26\degree\,24\arcm\,59\arcs$ (J2000.0) by Govoni et
al. \cite{gov01b}]. We also consider the bright galaxy ID~81
(hereafter BCM--D, see Sect.~\ref{sec:2D}) located in the middle of
the above three galaxies.

All the galaxies assigned to the A115 peak are analysed in a second
step by applying the ``shifting gapper'' technique by Fadda et
al. (\cite{fad96}), which uses the combination of position and
velocity information.  This procedure rejects galaxies that are too
far in velocity from the main body of galaxies and within a fixed bin
that shifts along the distance from the cluster centre.  The procedure
is iterated until the number of cluster members converges to a stable
value.  Following Fadda et al. (\cite{fad96}) we use a gap of $1000$
\ks -- in the cluster rest--frame -- and a bin of 0.6 \hh, or large
enough to include 15 galaxies. In the case of the A115 complex, which
is not an individual system, we fix alternatively BCM--A and --B as
cluster centres.  Fig.~\ref{figvd} shows the results. The very
high velocity galaxy (ID~91) is a clear interloper far more than
3000 \ks from other galaxies.  The other two high velocity galaxies
(IDs~96 and 100), that are close enough in 2D and far from the high
velocity BCM--A galaxy (see Fig.~\ref{figimage}), are rejected in both
our analyses. The conclusion about the low velocity tail is less clear
also because few of these galaxies lie in the eastern region of the
cluster and might be associated to the C clump (or to the D clump, see
Sect.~\ref{sec:2D}). We decide to define 85 likely cluster members
rejecting the three highest velocity galaxies (basic sample). In some
analyses we also consider an alternative sample of 80 galaxies --
Sample2 -- also rejecting the low velocity galaxies indicated in
Fig.~\ref{figvd} (the crosses in the middle panel).

By applying the bi-weight estimator to cluster members (Beers et
al. \cite{bee90}), we compute a mean cluster redshift of
$\left<z\right>=0.1929\pm$ 0.0005, i.e.
$\left<\rm{v}\right>=57817\pm$148 \kss.  We estimate the LOS velocity
dispersion, $\sigma_{\rm v}$, by using the bi-weight estimator and
applying the cosmological correction and the standard correction for
velocity errors (Danese et al. \cite{dan80}).  We obtain $\sigma_{\rm
v}=1362_{-108}^{+126}$ \kss, where errors are estimated through a
bootstrap technique. Consistent results are found for Sample2 (see
Table~\ref{tabv}).

\addtocounter{table}{+1}
\begin{table}
        \caption[]{Results of the kinematical analysis.}
         \label{tabv}
                $$
         \begin{array}{l r l l c}
            \hline
            \noalign{\smallskip}
            \hline
            \noalign{\smallskip}
\mathrm{Sample} & \mathrm{N_g} & \phantom{249}\mathrm{<v>}\phantom{249} & 
\phantom{24}\sigma_{\rm v}^{\mathrm{a}}\phantom{24}& \mathrm{BCM\ 
galaxies}\\
& &\phantom{249}\mathrm{km\ s^{-1}}\phantom{249} 
&\phantom{2}\mathrm{km\ s^{-1}}\phantom{24} & \\
            \hline
            \noalign{\smallskip}
 
\mathrm{Whole\ system}         &85 &57817\pm148 &1362_{-108}^{+126}& 
\mathrm{A,B,C,D}\\
\mathrm{Whole\ system\ Sample2}&80 &57956\pm132 
&1175_{-76}^{+95}&\mathrm{A,B,C}\\
\mathrm{WGAP1}        & 8 &54448\pm212 & 529_{-44}^{+124}&\mathrm{D}\\
\mathrm{WGAP2}        & 6 &55912\pm78  & 162_{-55}^{+39}& \\
\mathrm{DS-A}         &11 &59199\pm263 & 
816_{-103}^{+128}&\mathrm{A}\\
\mathrm{DS-C}          & 3 &56138\pm1001& - &\mathrm{C}  \\
\mathrm{KMM1}         &10 &54819\pm299 & 873_{-114}^{+225}&\mathrm{D} \\
\mathrm{KMM2}      &44 &57833\pm155 &1017_{-77}^{+109}&\mathrm{B,C} \\
\mathrm{KMM3}      &31 &58500\pm203 &1109_{-112}^{+199}&\mathrm{A} \\
\mathrm{CORE-A}       & 6 &59457\pm364 & 750_{-153}^{+337}&\mathrm{A}\\
\mathrm{CORE-B}       & 6 &57493\pm434 & 894_{-597}^{+290}&\mathrm{B}\\
\mathrm{Passive\ gals}       & 34 &57325\pm252 & 
1447_{-152}^{+247}&^{\mathrm{b}}\mathrm{C,D}\\
\mathrm{Active\ gals}       &14  &58950\pm411 & 
1462_{-440}^{+704}&\mathrm{A}\\
\mathrm{Balmer\ abs.\ gals}     & 9 &58988\pm668 & 1817_{-994}^{+834}&\\
\mathrm{Emission\ lines\ gals } & 5 &58558\pm527 & 898_{-218}^{+363}&\mathrm{A}\\
              \noalign{\smallskip}
            \hline
            \noalign{\smallskip}
            \hline
         \end{array}
$$
\begin{list}{}{}  
\item[$^{\mathrm{a}}$] We use the biweigth and the gapper estimators by
Beers et al. (1990) for samples with $\mathrm{N_g}\ge$ 15 and with
$\mathrm{N_g}<15$ galaxies, respectively (see also Girardi et
al. \cite{gir93}). 
\item[$^{\mathrm{b}}$] Notice that BCM-B resembles characteristics of 
a passive galaxy, but it does not appears in our classification
since its spectra has $\cal R$ slightly below the threshold value of 7. 
\end{list}
         \end{table}

Hereafter, for practical reasons, we consider as centre of the whole
A115 complex the position of the bi-weight centre obtained using
bi-weight mean estimators for R.A. and Dec. separately
[R.A.=$00^{\mathrm{h}}56^{\mathrm{m}}01\dotsec31$, Dec.=$+26\degree
22\arcmm 26\dotarcs6$ (J2000.0)].

\subsection{Velocity structure}
\label{sec:vel}

We analyse the velocity distribution to look for possible deviations
from Gaussianity that could provide important signatures of complex
dynamics. For the following tests the null hypothesis is that the
velocity distribution is a single Gaussian.

We estimate three shape estimators, i.e. the kurtosis, the skewness,
and the scaled tail index (see, e.g., Beers et al.~\cite{bee91}).  The
value of the skewness (-0.462) shows evidence that the velocity
distribution differs from a Gaussian at the $95-99$\% c.l. (see
Table~2 of Bird \& Beers~\cite{bir93}). Moreover, the W--test (Shapiro
\& Wilk \cite{sha65}) marginally rejects the null hypothesis of a
Gaussian parent distribution at the 92\% c.l..

\begin{figure}
\centering 
\resizebox{\hsize}{!}{\includegraphics{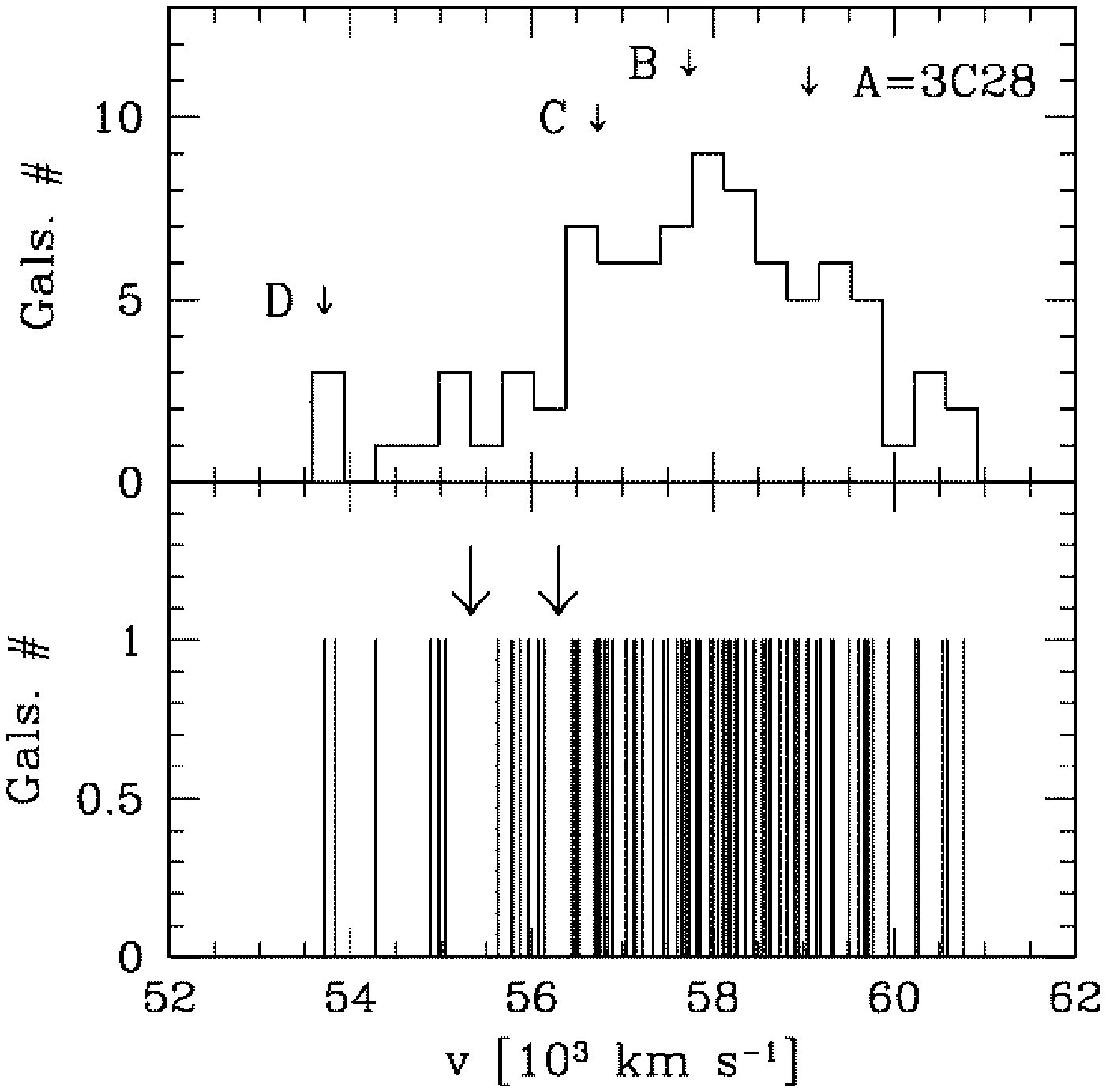}}
\caption{{\em Top panel}: velocity distribution of the 85 fiducial cluster
members. Arrows correspond to the bright galaxies BCM--A, --B, --C and --D).
{\em Bottom panel}: stripe density plot where the arrows indicate
the positions of the significant gaps. The gap at the lower velocity has 
a normalized size =2.41, the other  =2.25).}
\label{figstrip}
\end{figure}

Then we investigate the presence of gaps in the distribution.  A
weighted gap in the space of the ordered velocities is defined as the
difference between two contiguous velocities, weighted by the location
of these velocities with respect to the middle of the data. We obtain
values for these gaps relative to their average size, precisely the
midmean of the weighted--gap distribution. We look for normalized gaps
larger than 2.25 since in random draws of a Gaussian distribution they
arise at most in about 3\% of the cases, independent of the sample
size (Wainer and Schacht~\cite{wai78}; see also Beers et
al.~\cite{bee91}). Two significant gaps in the ordered velocity
dataset are detected individuating two groups in the low velocity tail
of the velocity distribution, of eight and six galaxies (see WGAP1 and
WGAP2 in Table~\ref{tabv} and Fig.~\ref{figstrip}).

We use the results of the gap analysis to determine the first guess
when using the Kaye's mixture model (KMM) to find a possible group
partition of the velocity distribution (as implemented by Ashman et
al. \cite{ash94}). The KMM algorithm fits an user--specified number of
Gaussian distributions to a dataset and assesses the improvement of
that fit over a single Gaussian. In addition, it provides the
maximum--likelihood estimate of the unknown n--mode Gaussians and an
assignment of objects into groups.  We do not find any two-- or
three--groups partition which is a significant better descriptor of the
velocity distribution with respect to a single Gaussian.

\subsection{2D galaxy distribution}
\label{sec:2D}

\begin{figure}
\centering
\includegraphics[width=9cm]{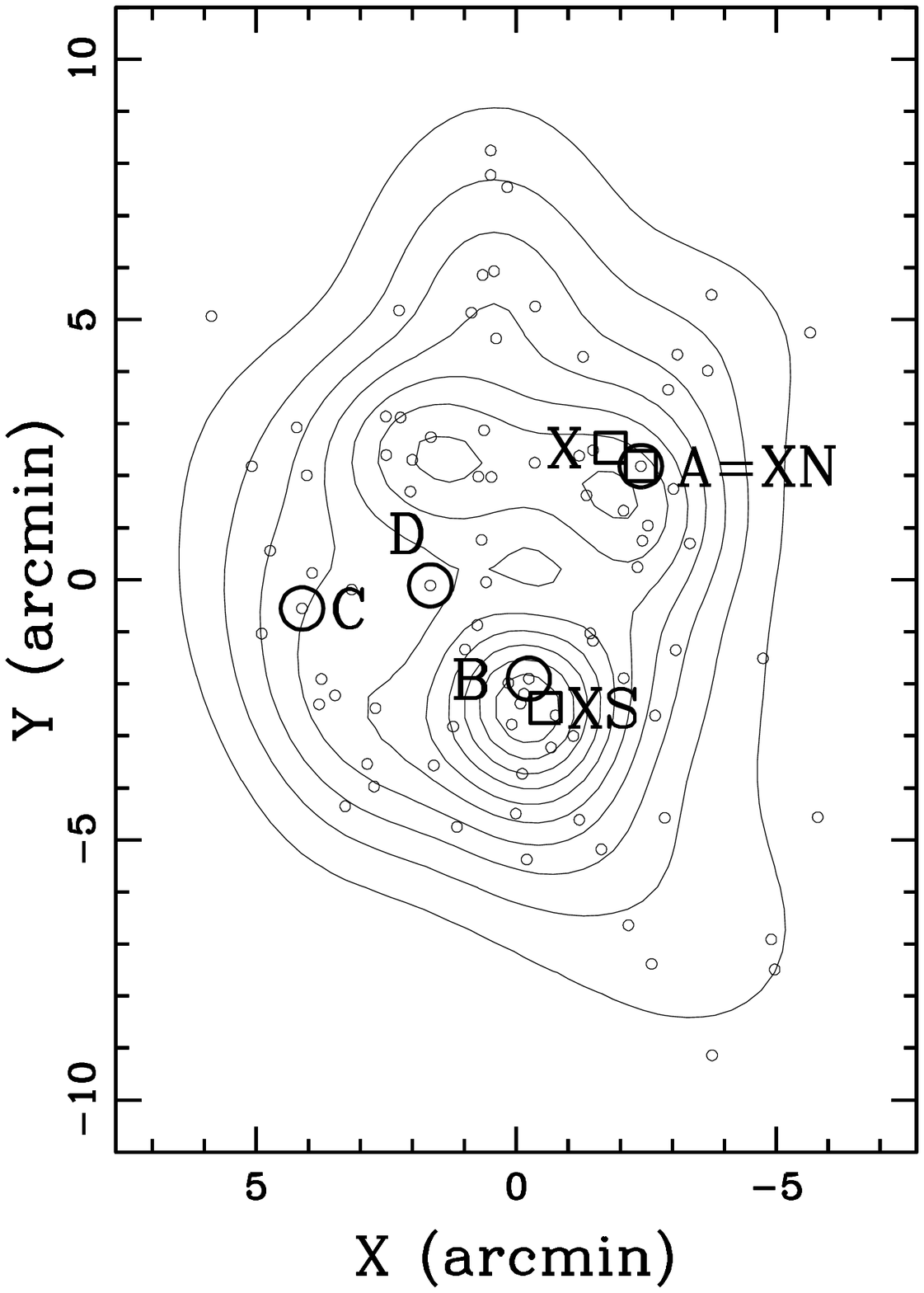}
\caption
{Spatial distribution on the sky of spectroscopically confirmed
cluster members and the relative isodensity contour map. The three
brightest galaxies BCM--A --B, and --C corresponding to Beers
et al. (\cite{bee83}) groups, a fourth bright
galaxy (BCM--D) and the X--ray peaks are indicated by large squares,
too.  The plot is centred on the cluster centre defined as the
bi-weight centre (see text).}
\label{figk2z}
\end{figure}

When applying the DEDICA method to the 2D distribution of A115 galaxy
members we find three significant peaks. The position of the highest
peak is close to the location of the BCM--B and of the southern peak of
X--ray emission (XS in Fig.\ref{figk2z}). Another peak is close to the
location of the BCM--A, of the northern peak of X--ray emission (XN)
and of the X--ray centroid when masking the X--ray emission of the
point source (X, Govoni et al. \cite{gov01b}). When dividing the
sample in bright and faint galaxies -- using the median magnitude
value $R$=18.32 -- we find that the 2D distributions of the two samples
are different at the 98.5\% c.l.  according to the two--dimensional
Kolmogorov--Smirnov test (hereafter 2DKS--test; see Fasano \&
Franceschini \cite{fas87}, as implemented by Press et
al. \cite{pre92}).  In fact, the faint galaxies sample shows both
peaks A and B, while the bright galaxies sample only shows the peak B.
 
\begin{figure}
\centering
\includegraphics[width=8cm]{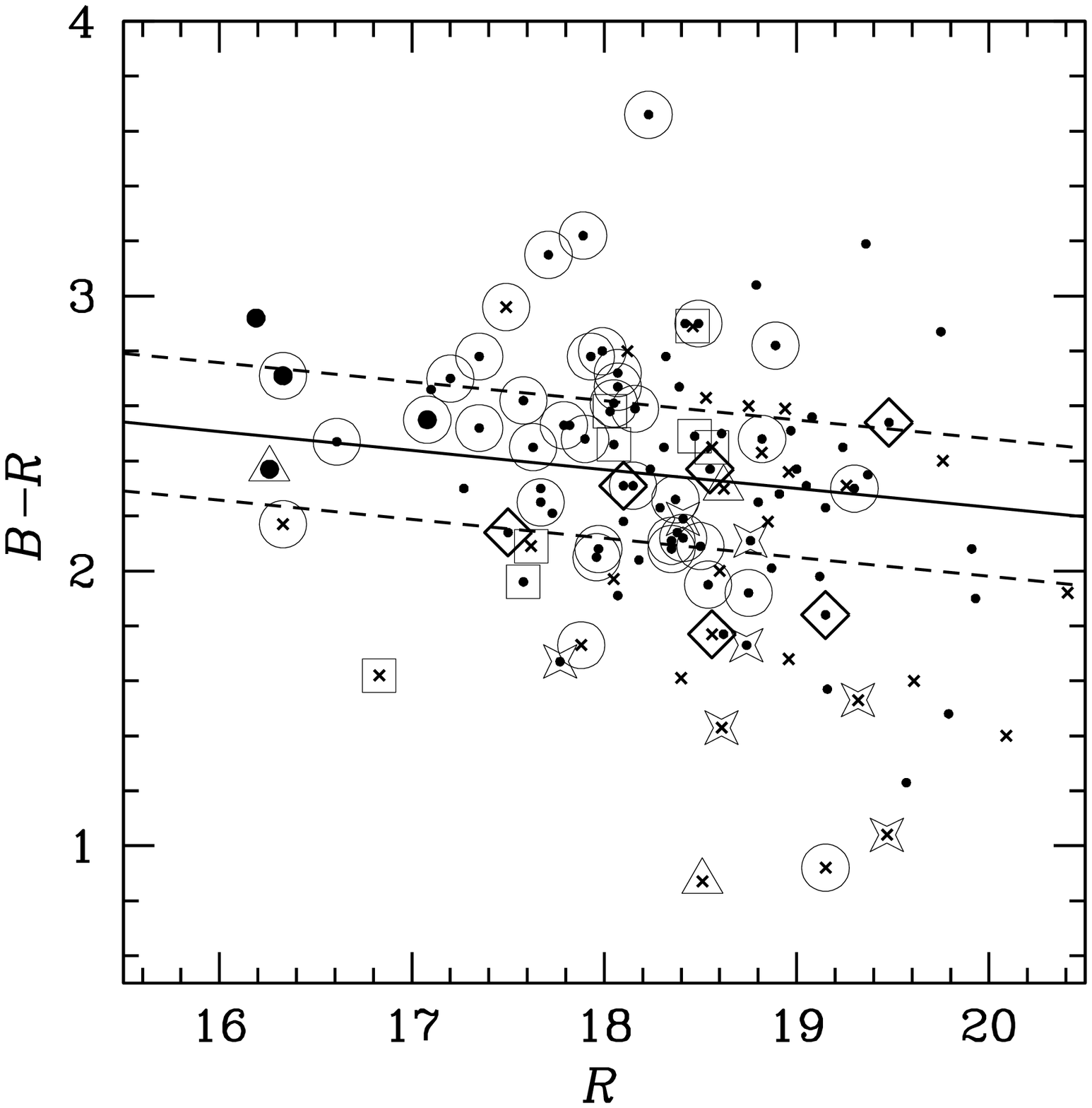}
\caption{$B$--$R$ vs. $R$ diagram for galaxies with available spectroscopy is shown
by small dots and crosses (cluster and field members, respectively).
Large solid dots indicate luminous galaxies BCM--A, --B, --C and
--D.  The solid line gives the best--fit colour--magnitude relation as
determined on member galaxies; the dashed lines are drawn at $\pm$0.25
mag from the CMR.  Large open symbols indicates classified galaxies:
circles (``k''), squares (``k+a''), rombs (``a+k''), triangles
(``e(c)''), and stars (``e(a)''/``e(b)'').  }
\label{figcmspettri}
\end{figure}

\begin{figure}
\centering
\includegraphics[width=8cm]{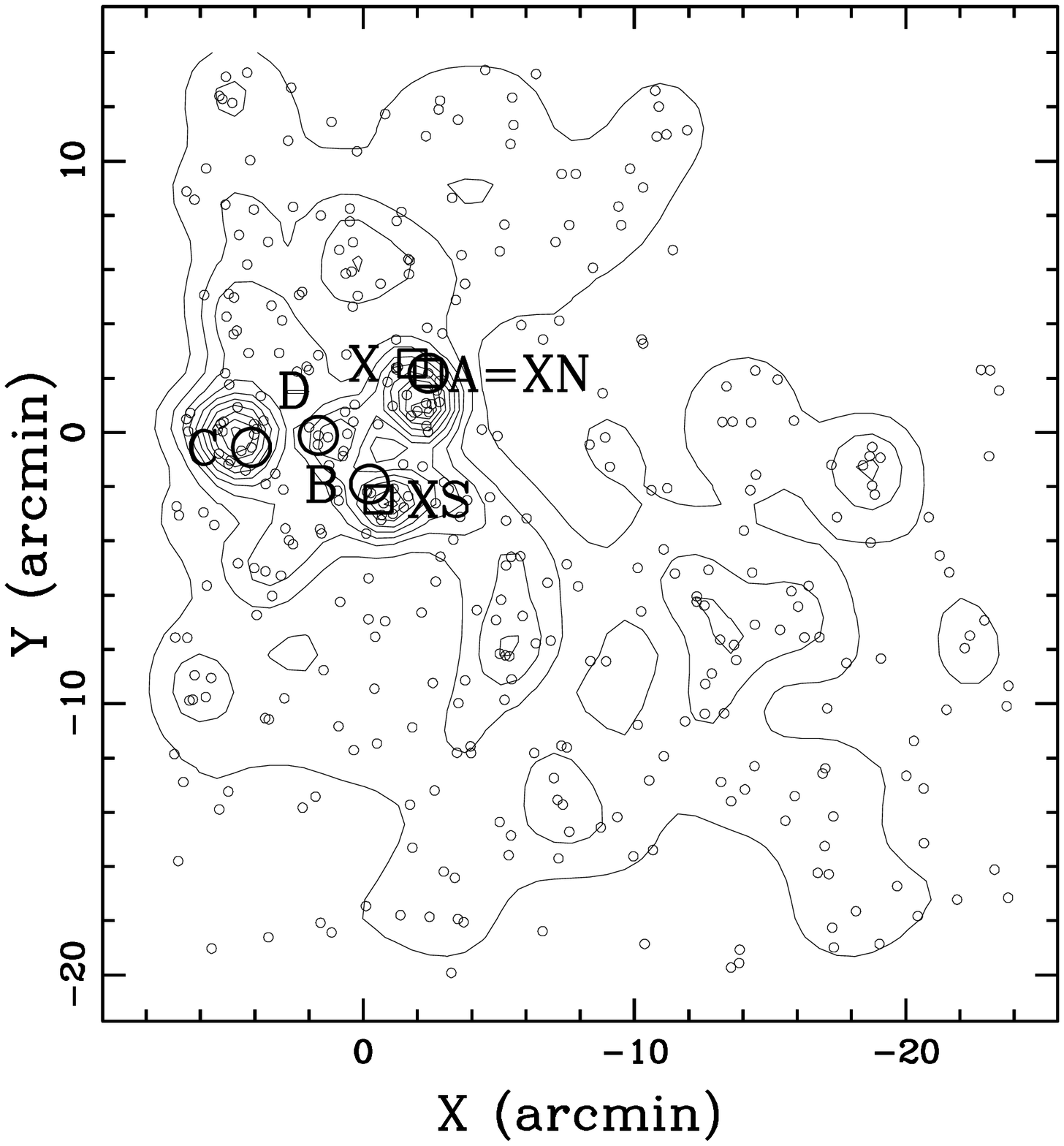}
\caption
{Spatial distribution on the sky and relative isodensity contour map
of 369 likely cluster members (according to the colour--magnitude
relation) with $R\le 21$, obtained with the DEDICA method.  The three
brightest galaxies BCM--A --B, and --C are indicated by large circles and
the X--ray peaks are indicated by large squares. It is also shown the
position of the luminous galaxy BCM--D. The plot is centred on the
cluster centre.}
\label{figk2A}
\end{figure}

Our spectroscopic data do not cover the entire cluster field and
suffer from magnitude incompleteness. To overcome these limits we
recover our photometric catalogue selecting likely members on the
basis of the colour--magnitude relation (hereafter CMR), which
indicates the early--type galaxy locus.  To determine CMR we fix the
slope according to L\'opez--Cruz et al. (\cite{lop04}, see their
Fig.~3) and apply the two--sigma--clipping fitting procedure to the
cluster members obtaining $B$--$R=3.611-0.069\times R$ (see
Fig.~\ref{figcmspettri}).  Out of our photometric catalogue we
consider galaxies (objects with SExtractor stellar index $\le 0.9$)
lying within 0.25 mag of the CMR.  To avoid contamination by field
galaxies we do not show results for galaxies fainter than 21 mag (in
$R$--band).  The contour map for 369 likely cluster members having
$R\le 21$ shows again the two peaks in correspondence of BCM--A and
BCM--B and also a peak in correspondence of BCM--C. Finally, a less
dense peak lies in the middle of the above three peaks and corresponds
to the luminous galaxy BCM--D at a very low velocity (see
Fig.~\ref{figk2A}).  Similar results are found analysing the 268
galaxies with $R\le 20$.  The analysis of 136 galaxies with $R\le 19$
shows as very significant only the southern peak close to BCM--B in
agreement with the results coming from the spectroscopic sample (see
above).

\subsection{Position--velocity correlations}
\label{sec:posvel}

The existence of correlations between positions and velocities of
cluster galaxies is a footprint of real substructures.  Here we use
different approaches to analyse the structure of A115 combining
velocity and position information.

\begin{figure}
\centering
\resizebox{\hsize}{!}{\includegraphics{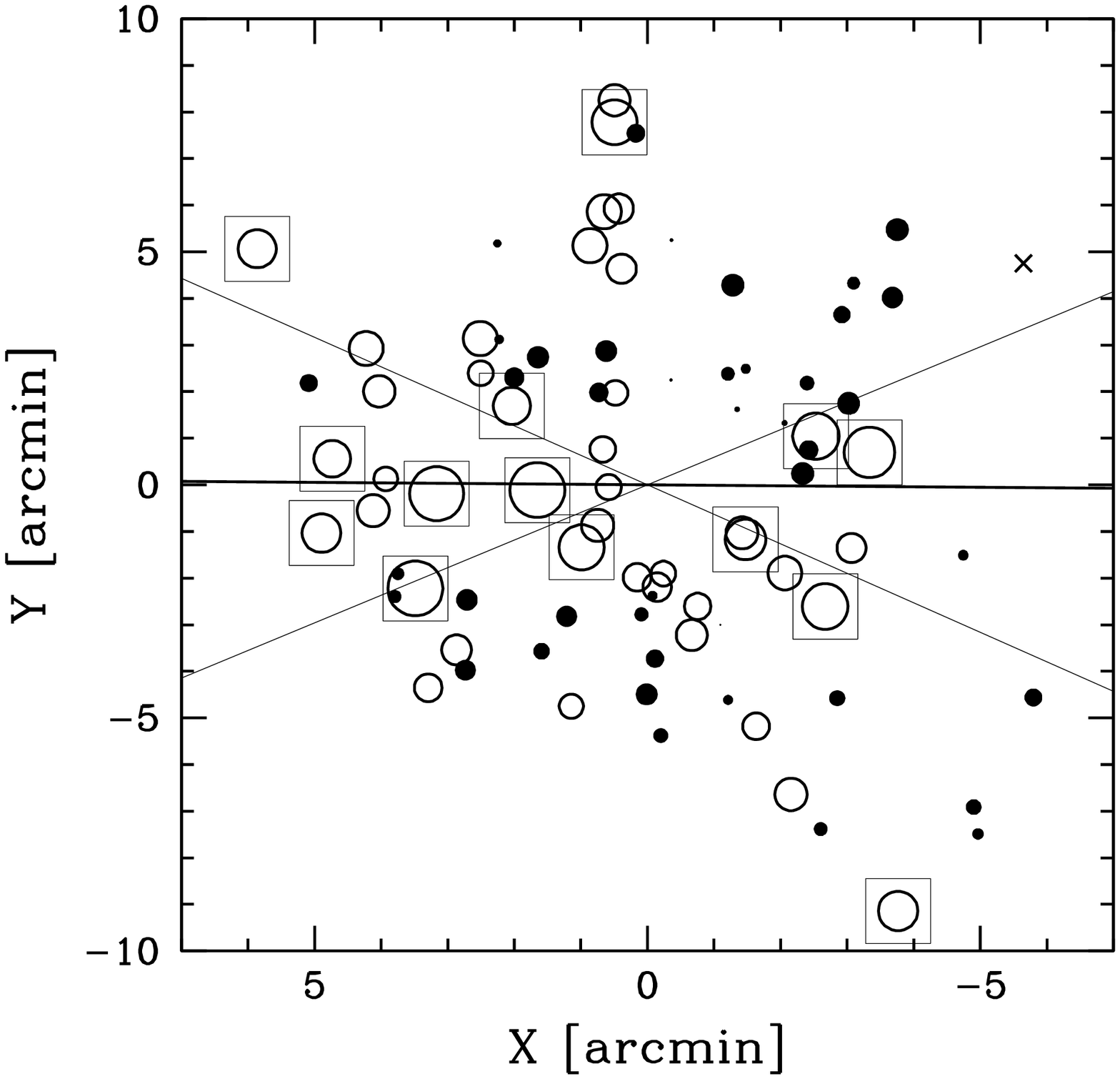}}
\caption
{Spatial distribution on the sky of 85 cluster members.  Open and
solid circles indicate low and high velocity galaxies, the cross the
galaxy with median velocity.  The larger the symbol, the smaller is
the radial velocity.  
The
solid and faint lines indicate the position angle of the cluster
gradient and relative errors, respectively. The faint big squares
indicate the 14 galaxies belonging to WGAP1 and WGAP2 (see
Sect.~\ref{sec:vel}).
The plot is centred on the cluster centre.  
}
\label{figgrad}
\end{figure}

The cluster velocity field may be influenced by the presence of
internal substructures.  To investigate the velocity field of the A115
complex we divide galaxies in a low and a high velocity samples by
using the median cluster velocity and check the difference between the
two distributions of galaxy positions. Fig.~\ref{figgrad} shows that
low and high velocity galaxies are segregated roughly in the E--W
direction.  The two distributions are different at the 99.5\% c.l.
according to the 2DKS--test.  In order to estimate the direction of
the velocity gradient we perform a multiple linear regression fit to
the observed velocities with respect to the galaxy positions in the
plane of the sky (see also den Hartog \& Katgert \cite{den96}; Girardi
et al. \cite{gir96}). We find a position angle on the celestial sphere
of $PA=269_{-18}^{+19}$ degrees (measured counter--clock--wise from
north), i.e. higher velocity galaxies lie in the western region of the
cluster (see Fig.~\ref{figgrad}). To assess the significance of this
velocity gradient we perform 1000 Monte Carlo simulations by randomly
shuffling the galaxy velocities and for each simulation we determine
the coefficient of multiple determination ($RC^2$, see e.g., NAG
Fortran Workstation Handbook \cite{nag86}).  We define the
significance of the velocity gradient as the fraction of times in
which the $RC^2$ of the simulated data is smaller than the observed
$RC^2$.  We find that the velocity gradient is marginally significant
at the 90\% c.l..  Similar results are obtained for the Sample2
($PA=267_{-17}^{+18}$ degrees at the 91\% c.l.).

\begin{figure}
\centering 
\resizebox{\hsize}{!}{\includegraphics{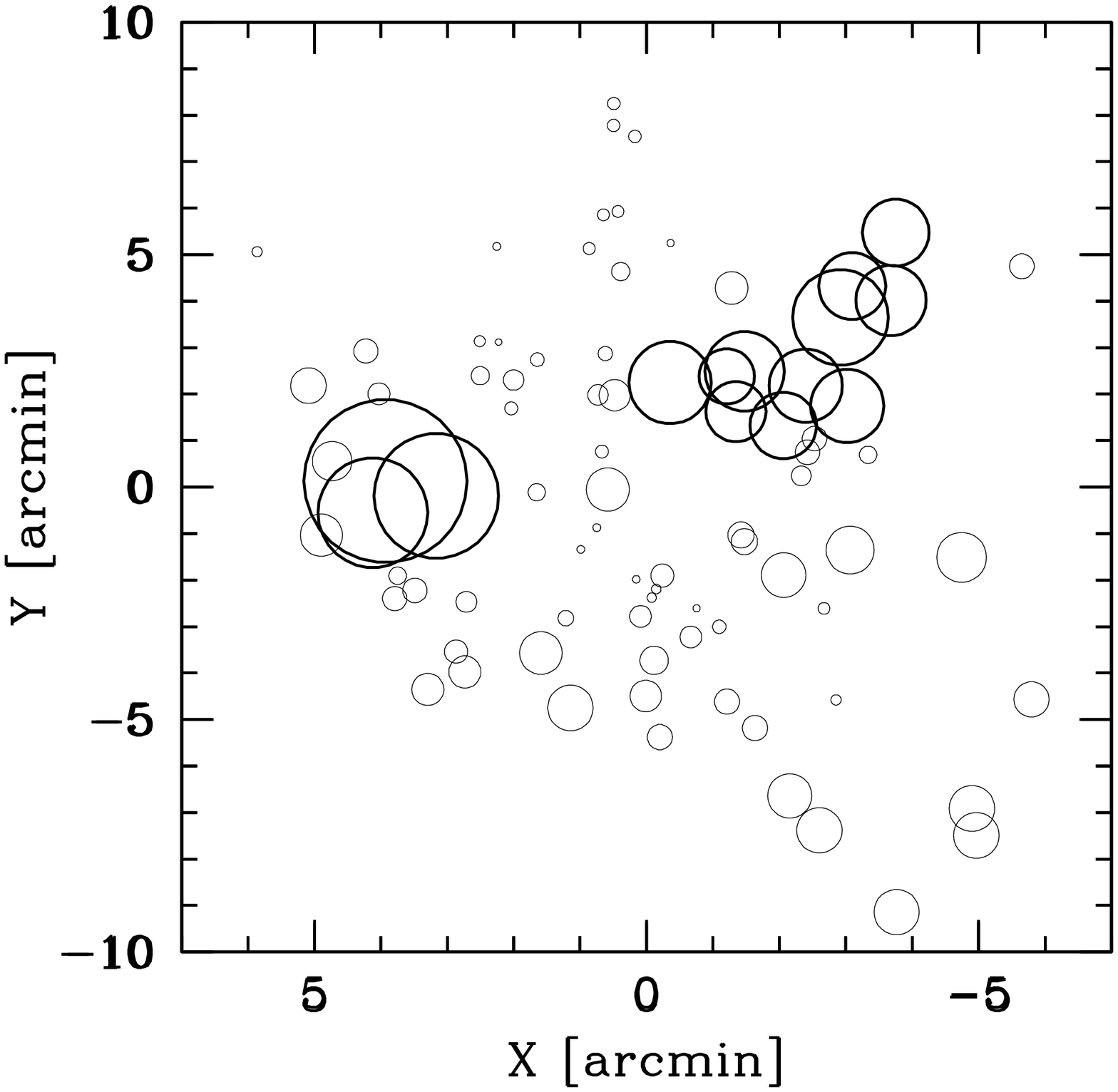}}
\caption
{Spatial distribution of cluster members, each marked by a circle: the
larger the circle, the larger is the deviation $\delta_i$ of the local
parameters from the global cluster parameters, i.e. there is more
evidence for substructure (according to the Dressler \& Schectman
test, see text).  The boldface circles indicate those with $\delta_i
\ge 2.2$ (see text).
The plot is centred on the cluster centre.  
}
\label{figds}
\end{figure}

We combine galaxy velocity and position information to compute the
$\Delta$--statistics devised by Dressler \& Schectman (\cite{dre88}).
This test is sensitive to spatially compact subsystems that have
either an average velocity that differs from the cluster mean, or a
velocity dispersion that differs from the global one, or both.  We
find $\Delta=121$ for the value of the parameter which gives the
cumulative deviation.  This value is an indication of substructure,
significant at the 96\% c.l., as assessed computing 1000 Monte Carlo
simulations, randomly shuffling the galaxy velocities.
Fig.~\ref{figds} shows the distribution on the sky of all galaxies,
each marked by a circle: the larger the circle, the larger the
deviation $\delta_i$ of the local parameters from the global cluster
parameters, i.e. the higher the evidence for substructure.  This
figure provides information on the positions of substructures: one in
the eastern region corresponding to the clump C and one in the
northern region corresponding to the clump A.  Similar results
obtained for Sample2 ($\Delta=110$ and a c.l. of 95\%), but the
eastern substructure is no longer so obvious in the relative plot.

\begin{figure}
\centering 
\resizebox{\hsize}{!}{\includegraphics{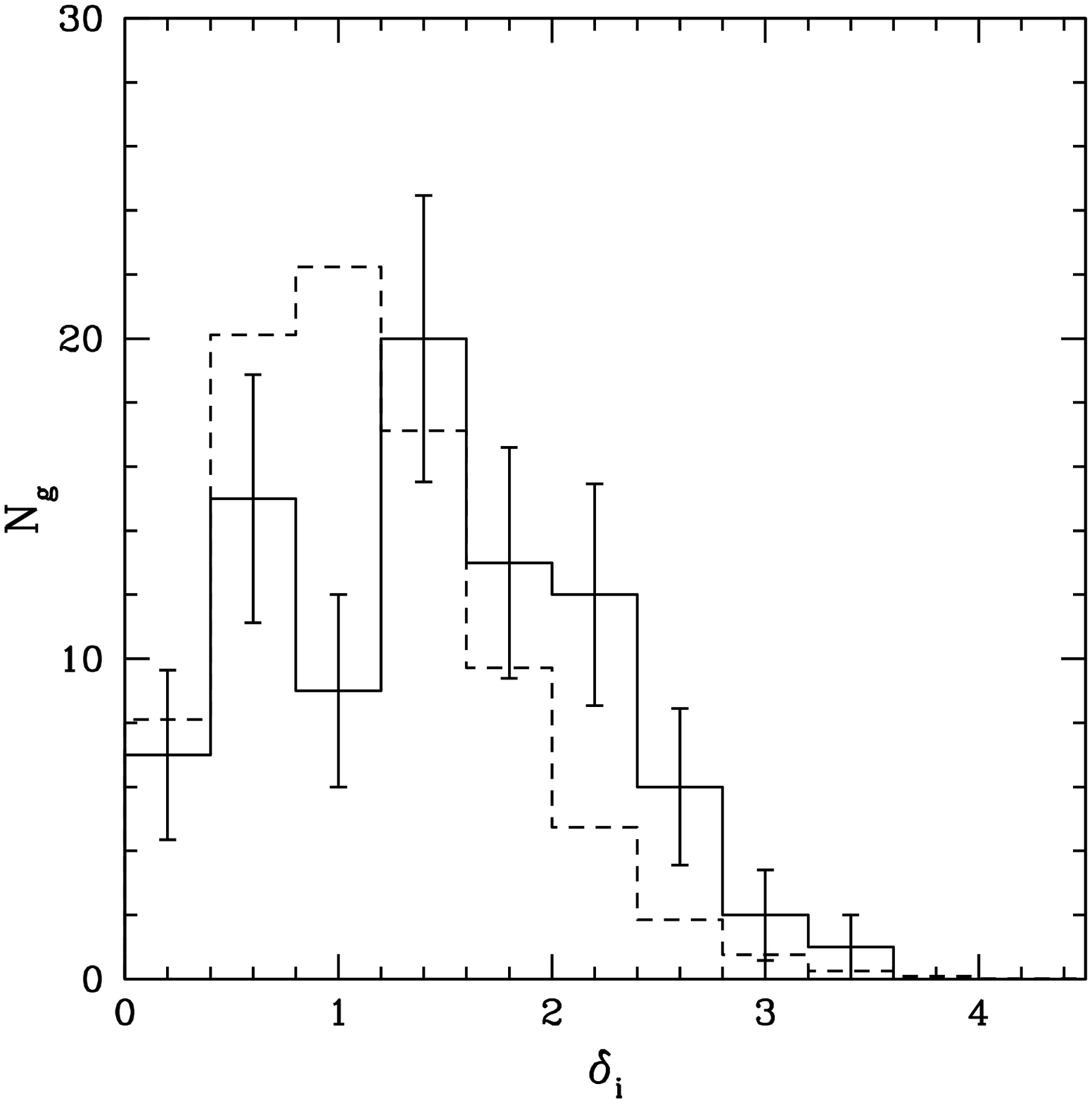}}
\caption
{The distribution of $\delta_i$ deviations of the Dressler--Schectman
analysis for the 85 member galaxies. The solid line represents the
observations, the dashed line the distribution for the galaxies of
simulated clusters, normalized to the observed number.}
\label{figdeltai}
\end{figure}

To obtain further information, we resort to the technique developed by
Biviano et al. (\cite{biv02}), who used the individual
$\delta_i$--values of the Dressler \& Schectman method. The critical
point is to determine the value of $\delta_i$ that optimally indicates
galaxies belonging to substructure. To this aim we consider the
$\delta_i$--values of all 1000 Monte Carlo simulations already used to
determine the significance of the substructure (see above).  The
resulting distribution of $\delta_i$ is compared to the observed one
finding a difference of $P>99.99$\% c.l. according to the KS--test.
The ``simulated'' distribution is normalized to produce the observed
number of galaxies and compared to the observed distribution in
Fig.~\ref{figdeltai}: the latter shows a tail at large values. The
tail with $\delta \gtrsim 2$ is populated by galaxies that presumably
are in substructures.  For the selection of galaxies within
substructures we choose the value of $\delta_{\rm lim}=2.2$, since the
galaxies with $\delta_i>\delta_{\rm lim}$ are well separated in the sky
(see Fig.~\ref{figds}) and assign three and eleven galaxies to the
clumps C and A, respectively (hereafter DS--C and DS--A).  The northern
structure is populated by high velocity galaxies, while the poor
statistics prevent us to obtain firm conclusions about the eastern
structure (see Table~\ref{tabv}).

\begin{figure}
\centering 
\resizebox{\hsize}{!}{\includegraphics{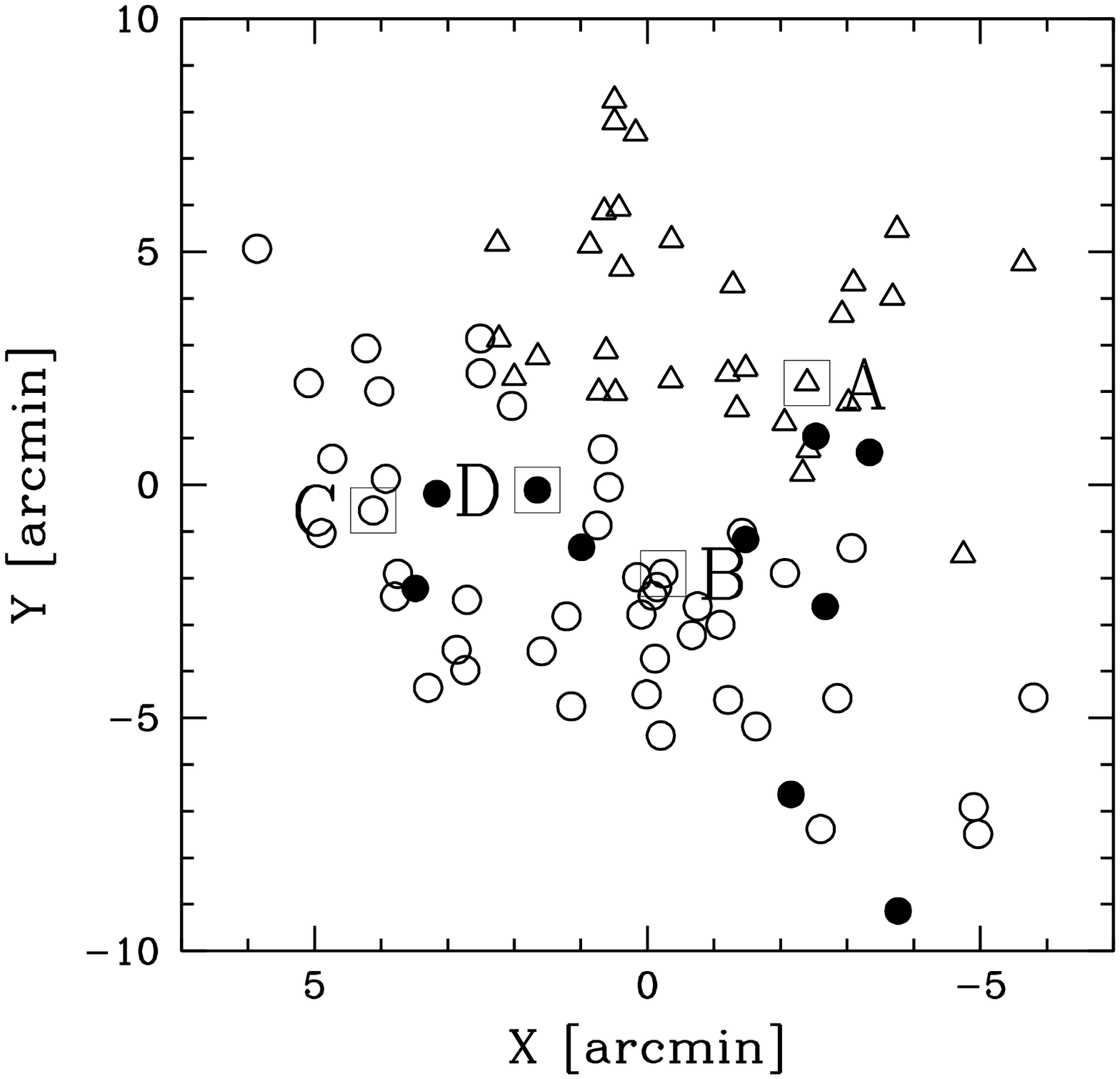}}
\caption
{Spatial distribution on the sky of the 85 member galaxies. Solid
circles, open circles and triangles indicate KMM1, KMM2, and KMM3.
The large faint squares
indicate the position of the brightest cluster members BCM--A, --B, --C
and --D.
The plot is centred on the cluster centre.  
}
\label{figkmm}
\end{figure}

The Gaussian model for the 2D galaxy distribution is poorly supported
by theoretical and/or empirical arguments and, however, our galaxy
catalogue is not complete down to a magnitude limit.  However, since
the 3D diagnostics is in general the most sensitive indicators of the
presence of substructure (e.g., Pinkney et al. \cite{pin96}), we apply
the 3D version of the KMM software using simultaneously galaxy
velocity and positions. We use the galaxy assignment given by
Dressler--Schectman method to determine the first guess when fitting
three groups.  The algorithm fits a three--groups partition at the
97\% c.l. according to the likelihood ratio test (hereafter KMM1,
KMM2, KMM3 groups from low to high mean velocities).  The results for
the three groups are shown in Table~\ref{tabv} and Fig.~\ref{figkmm}.
KMM1 group is sparse in the sky, but well distinct in velocity from
the other two groups.  Several of its galaxies were already detected
by the weighted gap analysis as belonging to WGAP1 and WGAP2 (see
Sect.~\ref{sec:vel}). Moreover, KMM1 contains BCM--D.  KMM2 and
KMM3 groups are well distinct in the sky. KMM2 contains both
BCM--B and BCM--C, while KMM3 contains BCM--A.

\subsection{Kinematics of A115N and A115S}
\label{sec:kin}

\begin{figure}
\centering 
\resizebox{\hsize}{!}{\includegraphics{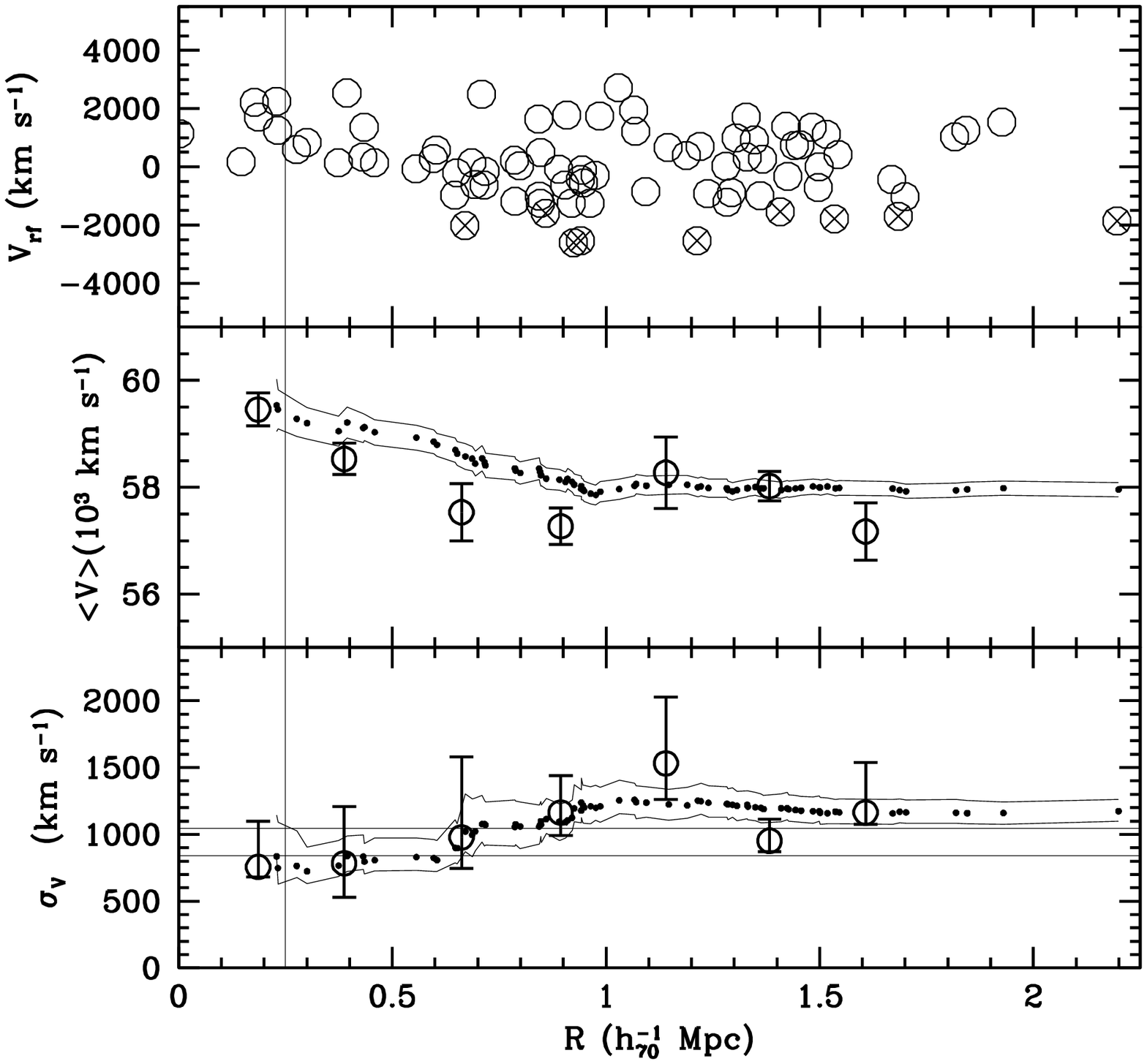}}
\caption
{Kinematical profiles of the northern subcluster A115N obtained
assuming BCM--A as centre.  The vertical line indicates the region
likely not contaminated from other galaxy clumps (see text).  {\em Top
panel}: rest--frame velocity vs. projected distance from the
subcluster centre (BCM--A): crosses indicate the galaxies belonging to
WGAP1 and WGAP2.  Differential (big circles) and integral (small
points) mean velocity and LOS velocity--dispersion profiles are shown
in {\em middle and bottom panels}, respectively. For the differential
profiles we plot the values for seven annuli from the centre of the
subcluster, each of 0.25 \hh.  For the integral profiles, the mean and
dispersion at a given (projected) radius from the subcluster--centre
is estimated by considering all galaxies within that radius -- the
first value computed on the five galaxies closest to the centre. The
error bands at the $68\%$ c.l. are also shown.  In the bottom panel,
the horizontal line represents the range of X--ray temperatures as
given in the literature for A115N (see Sect.~\ref{sec:int})
transformed in $\sigma_{\rm v}$ assuming the density--energy
equipartition between gas and galaxies, i.e.  $\beta_{\rm spec}=1$
(see text).  }
\label{figprofxa}
\end{figure}

\begin{figure}
\centering 
\resizebox{\hsize}{!}{\includegraphics{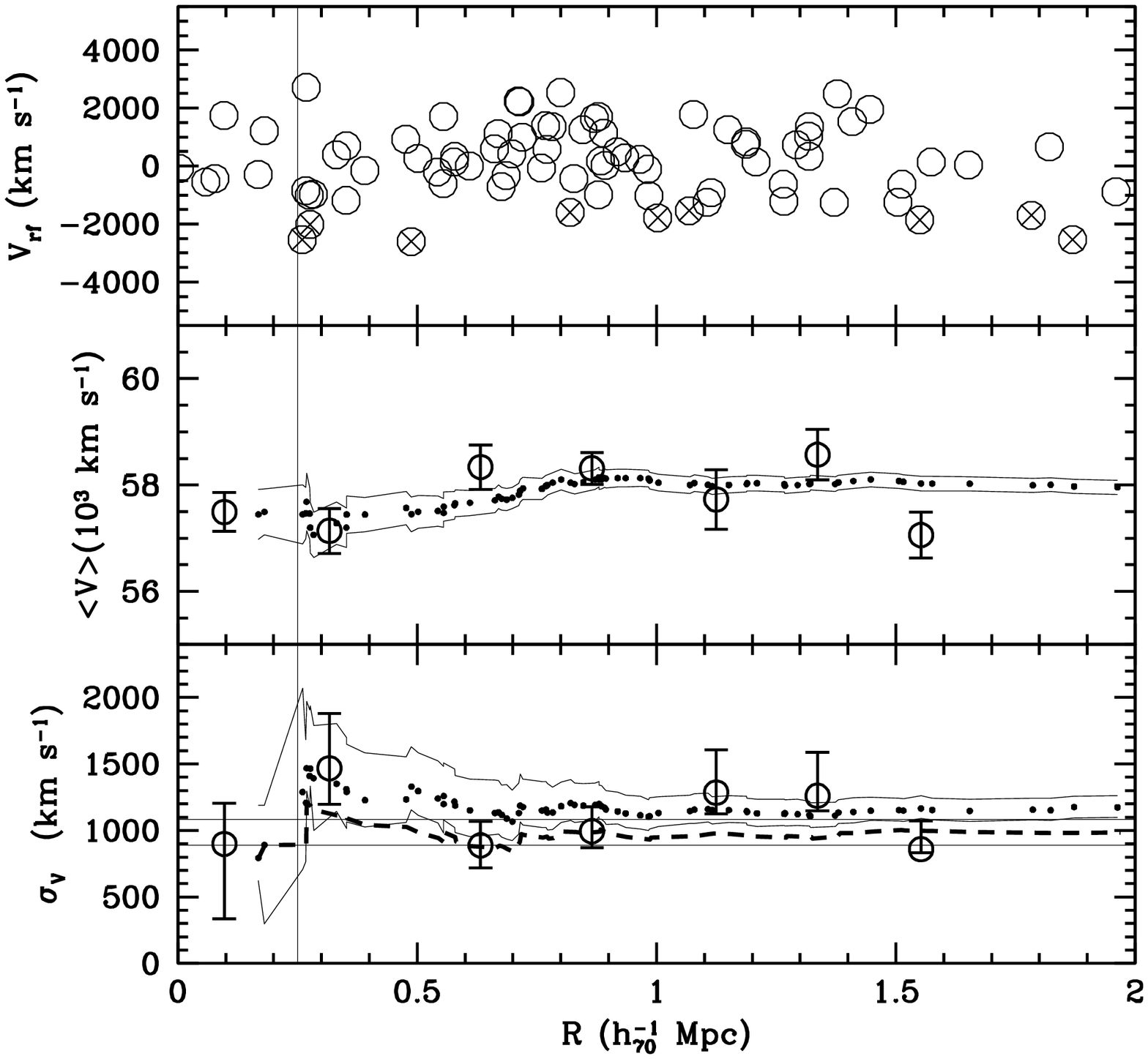}}
\caption
{ The same as in Fig.~\ref{figprofxa}, but referring to the southern
subcluster A115S centred around BCM--B. The dashed line in the bottom
panel gives the integral velocity dispersion profile when rejecting
the low velocity galaxies of WGAP1 and WGAP2.  In the bottom panel,
the horizontal line represents the range of X--ray temperatures as
given in the literature for A115S (see Sect.~\ref{sec:int})
transformed in $\sigma_{\rm v}$ assuming the density--energy
equipartition between gas and galaxies, i.e.  $\beta_{\rm spec}=1$
(see text).  }
\label{figprofxb}
\end{figure}

The spatial agreement between the two brightest cluster members
(BCM--A and --B) and the peaks of X--ray emission as well as the high
density of galaxies around BCM--A and --B prompts us to analyse the
profiles of the mean velocity and the velocity dispersion of galaxy
systems surrounding these two galaxies (see Figs.~\ref{figprofxa} and
\ref{figprofxb}, respectively).  This allows an independent analysis
of the individual galaxy clumps. Although an increase in the
velocity--dispersion profile in the cluster central regions might be
due to dynamical friction and galaxy merging (e.g., Menci \&
Fusco--Femiano \cite{men96}; Girardi et al. \cite{gir98}; Biviano \&
Katgert \cite{biv04}), in the case of A115 it might be simply induced
by the contamination of the galaxies of a secondary clump (e.g.,
Girardi et al. \cite{gir96}; Girardi et al. \cite{gir05}). The latter
hypothesis can be investigated by looking at the behaviour of the mean
velocity profile.

Since from the above sections we know the presence of one or likely a
few low velocity groups, we analyse the Sample2 to avoid, at least
partially, the possible contamination.  We also consider the results
obtained rejecting all galaxies belonging to low velocity WGAP1 and
WGAP2.  Figs.~\ref{figprofxa} and \ref{figprofxb} show
velocity--dispersion and mean--velocity profiles, as well as regions
not likely to be contaminated by other galaxy systems and thus
reliable for kinematical analysis. Detailed results of this analysis
are included in Table~\ref{tabv} where the ``uncontaminated'' galaxy
systems are named as CORE--A and CORE--B.

Fig.~\ref{figprofxa} shows how the integral velocity--dispersion
slightly increases with the distance from the BCM--A.  Simultaneously,
the mean velocity shows a continuous decline from high values $\sim
59000$ \ks suggesting a strong contamination of galaxies from
structures connected with BCM--B, --C and --D, all having lower mean
velocities. In a conservative view we consider the likely
uncontaminated region within 0.25 \hh, where we find $\sigma_{\rm
v}\simeq 750$ \kss. Fig.~\ref{figprofxb} shows an enough robust mean
velocity and a sharp increase of the integral velocity--dispersion
with the distance from the BCM--B up to a peak value $\sim 1450$ \ks at
$\sim 0.3$ \hh. We interpret these features as the contamination of
the structures connected with BCM--A, --C and --D with higher and lower
mean velocities.  Their combination does not affect the mean velocity
but strongly increases the velocity dispersion.  In fact, the peak
value of the velocity dispersion goes down from 1450 to 1200 \ks when
rejecting low velocity galaxies of WGAP1 and WGAP2. We then consider
the likely uncontaminated region within 0.25 \hh, where we find
$\sigma_{\rm v}\simeq 900$ \kss.

\section{3D structure and virial mass of A115}
\label{sec:mass}

On the basis of the above section we conclude that A115 is formed by
two subclusters well distinct in the sky and centred around BCM--A
and --B, hereafter A115N and A115S, with the addition of several low
velocity galaxies, above all in the eastern cluster region, likely
organised in small groups (see Fig.~\ref{figgrad} and the PA of the
velocity gradient).

As for the low velocity groups we detect two 2D galaxy concentrations
around the two bright, low velocity galaxies BCM--C and --D.
Moreover, we find two possible groups in the low velocity tail of the
velocity distribution (WGAP1 containing BCM--D and WGAP2).  The
Dressler--Schectman test detects a substructure around the BCM--C,
too.  The presence of a group around BCM--D finds some support in the
faint X--ray emission shown there (see Gutierrez \& Krawczynski
\cite{gut05} and Fig.~\ref{figisofote}). That we find evidence of a
few groups rather than of an individual massive cluster agrees with
the absence of a third X--ray luminous peak.

As for A115N, we find comparable kinematical properties in the
analysis of DS--A and CORE--A, i.e.  $\left<\rm{v}\right> \gtrsim $
59000 \ks and $\sigma_{\rm v} \sim$ 800 \kss. Instead KMM3 is
contaminated by the presence of low velocity galaxies, e.g. IDs~10
and 19 close to BCM--A.  As for A115S, we find $\left<\rm{v}\right>
\sim $ 57500 \ks and $\sigma_{\rm v} \sim$ 900--1000 \ks (see KMM2 and
CORE--B). The velocity dispersion of the two subclusters are roughly
comparable to their average X--ray temperature as listed in the
literature and transformed in $\sigma_{\rm v}$ assuming the
density--energy equipartition between gas and galaxies, i.e.
$\beta_{\rm spec}=1$
\footnote{$\beta_{\rm spec}=\sigma_{\rm v}^2/(kT/\mu m_{\rm p})$ with
$\mu=0.58$ the mean molecular weight and $m_{\rm p}$ the proton mass.}
(see Figs.~\ref{figprofxa} and \ref{figprofxb}).  The two subclusters
differ by $\sim 2000$ \ks in the LOS velocity, i.e. almost three times
more than what found by Beers et al. (\cite{bee83}) with only 19
cluster members.  In agreement with Beers et al. (\cite{bee83}), we
find that A115S is dynamically somewhat more important than A115N,
while the contrary is found by X--ray data (Jones \& Forman
\cite{jon99}; White et al. \cite{whi97}; Gutierrez \& Krawczynski
\cite{gut05}; but see Shibata et al. \cite{shi99}). As already
suggested by Beers et al. (\cite{bee83}) the presence of 3C28 might
affects the X--ray results overestimating the X--ray luminosity and
the temperature of A115N.

Although A115 is likely in a phase of interaction (see the following
section), the two main galaxy subclusters are still well detectable
and roughly well coincident with the X--ray peaks.  Thus we assume
that each subcluster is in dynamical equilibrium to compute virial
quantities. Hereafter we assume $\sigma_{\rm v}=$ 750--850 \ks and
900--1000 \ks for A115N and A115S, respectively.

Following the prescriptions of Girardi \& Mezzetti (\cite{gir01}), we
assume for the radius of the quasi--virialized region R$_{\rm
vir,N}=0.17\times \sigma_{\rm v}/H(z) = 2.2$--2.5 \h and R$_{\rm
vir,S}=2.6$--2.9 \h for A115N and A115S, respectively -- see their
eq.~1 after introducing the scaling with $H(z)$ (see also eq.~ 8 of
Carlberg et al. \cite{car97} for R$_{200}$). Therefore, our
spectroscopic catalogue samples most of the virialized region.

One can compute the mass using the virial theorem (Limber \& Mathews
\cite{lim60}; see also, e.g., Girardi et al. \cite{gir98}) under the
assumption that mass follows galaxy distribution: $M=M_{\rm
svir}-\rm{SPT}$, where $M_{\rm svir}=3\pi/2 \cdot \sigma_{\rm v}^2{\rm
R}_{\rm PV}/G$ is the standard virial mass, R$_{\rm PV}$ a projected
radius (equal two times the harmonic radius), and SPT is the surface
pressure term correction (The \& White \cite{the86}).  The value of
R$_{\rm PV}$ depends on the size of the considered region, so that the
computed mass increases (but not linearly) with the increasing
considered region.  Since the two subclusters are interacting and we
do not cover the whole virialized region we use an alternative
estimate which was shown be good when R$_{\rm PV}$ is computed within
R$_{\rm vir}$ (see eq.~13 of Girardi et al. \cite{gir98}).  This
alternative estimate is based on the knowledge of the galaxy
distribution and, in particular, a galaxy King--like distribution with
parameters typical of nearby/medium--redshift clusters: a core radius
R$_{\rm core}=1/20\times {\rm R}_{\rm vir}$ and a slope--parameter
$\beta_{\rm fit}=0.8$, i.e. the volume galaxy density at large radii
as $r^{-3 \beta_{\rm fit}}=r^{-2.4}$ (Girardi \& Mezzetti
\cite{gir01}).  For the whole virialized region we obtain R$_{\rm
PV,N}=1.6$--1.8 \h and R$_{\rm PV,S}=$1.9--2.2 \hh.  As for the SPT
correction, we assume a 20\% computed combining data on many clusters
one (e.g., Carlberg et al. \cite{car97}; Girardi et al. \cite{gir98}).
This leads to virial masses $M_{\rm N}(<{\rm R}_{\rm
vir,N}=2.2-2.5\,\hhh)=8.0$--11.6 \mqua and $M_{\rm S}(<{\rm R}_{\rm
vir,S}=2.6-2.9\,\hhh)=13.8$ --18.9 \mqua for the two subclusters.

To compare our results with the estimate recovered from a X--ray
surface brightness deprojection analysis (White et al. \cite{whi97})
we assume that each subcluster is described by a King--like mass
distribution (see above) or, alternatively, a NFW profile where the
mass--dependent concentration parameter is taken from Navarro et
al. (\cite{nav97}) and rescaled by the factor $1+z$ (Bullock et
al. \cite{bul01}; Dolag et al. \cite{dol04}).  We obtain $M_{\rm N}
(<{\rm R}_{\rm out,N}=0.235\,\hhh)=(4.2-9.9)$ \mtre and $M_{\rm S}
(<{\rm R}_{\rm out,S}=0.157\,\hhh)=(2.5-5.5)$ \mtree.  The first
estimate is somewhat smaller and the second one is in agreement with
those found by White et al. (\cite{whi97}, see their Table~3 where
$M_{\rm grav, N} (<{\rm R}_{\rm out,N})=14.0$ \mtre and $M_{\rm grav, S}
(<{\rm R}_{\rm out,S})=5.0$ \mtre in our cosmology).

As for the mass of the whole system, the contribution of the low
velocity groups is of minor importance since they likely have low
velocity dispersion and the virial mass scales with $\sigma_{\rm
v}^3$. E.g., we estimate that $M\sim2.4$ \mqua for a $\sigma_{\rm v}\sim
500$ \ks group (see WGAP1 in Table~\ref{tabv}).  Considering the
possible presence of, at most, two of these $\sigma_{\rm v}\sim 500$
\ks groups, a reliable mass estimate of the whole system is then
M=2.2--3.5 \mquii, in agreement with rich clusters reported in the
literature (e.g., Girardi et al. \cite{gir98}; Girardi \& Mezzetti
\cite{gir01}). A smaller mass estimate is given by Govoni et al.
[\cite{gov01b}; $M (<\rm{R=2.3}\,\hhh)=0.6$--1.2 \mqui centred on the
X--ray centroid in A115N], but notice that it refers to a smaller
cluster region, i.e. likely excluding a large part of A115S.

\section{Merging scenario}
\label{sec:merg}

Since A115N and A115S are well detectable and optical and X--ray data
indicate a very similar location we are likely looking at the cluster
prior to merging. However, A115N and A115S subclusters are already
starting to interact, as suggested by several pieces of evidence: the
slight displacement between peaks of gas distribution and of galaxy
distribution (see our Fig.~\ref{figk2A}); the presence of an hot
region likely due to the interaction (Shibata et al. \cite{shi99};
Gutierrez \& Krawczynski \cite{gut05}).  Very noticeably, the largest
dimension of the radio relic is somewhat perpendicular to the axis
connecting A115N and A115S in agreement with being originated by shock
waves connected to the ongoing merger (e.g., Ensslin \& Br\"uggen
\cite{ens02}).

When the merging scenario is assumed to explain the presence of the
hot region located between A115N and A115S, a relative colliding
velocity is necessary to heat up the ICM temperature $kT$ from $\sim
4$ to $\sim 9$ keV (see Gutierrez \& Krawczynski
\cite{gut05}). Assuming that the two subclusters are to cause a
head--on collision and that their kinetic energies are completely
converted to thermal energy, the necessary value of the colliding
velocity is ${\rm v}_{\rm coll}=(3k \Delta T/\mu m_{\rm p})^{1/2}$
(see Shibata et al. \cite{shi99}). We find ${\rm v}_{\rm coll}\sim
1600$ \ks in good agreement with the observed relative LOS velocity in
the rest frame ${\rm V}_{\rm r}=1646$ \ks as recovered from CORE--A
and CORE--B in the cluster rest frame, i.e. ($<$v$>_{\rm
CORE-A}$-$<$v$>_{\rm CORE-B})/(1+z)$).

\subsection{Bimodal model}
\label{sec:bim}

Here we investigate the relative dynamics of A115N and A115S using
different analytic approaches which are based on an energy integral
formalism in the framework of locally flat spacetime and Newtonian
gravity (e.g., Beers et al. \cite{bee82}).  The values of the relevant
observable quantities for the two--clumps system are: the relative LOS
velocity in the rest frame, ${\rm V}_{\rm r}=1646$ \ks (as recovered
from CORE--A and CORE--B); the projected linear distance between the
two clumps, $D=0.89$ \h (as recovered from BCM--A and --B cluster rest
frame); the mass of the system obtained by adding the masses of the
two subclusters each within its virial radius, log$M_{\rm
sys}=15.4107_{-0.0734}^{+0.0734}$ (see Sect.~\ref{sec:mass}).

\begin{figure}
\centering
\resizebox{\hsize}{!}{\includegraphics{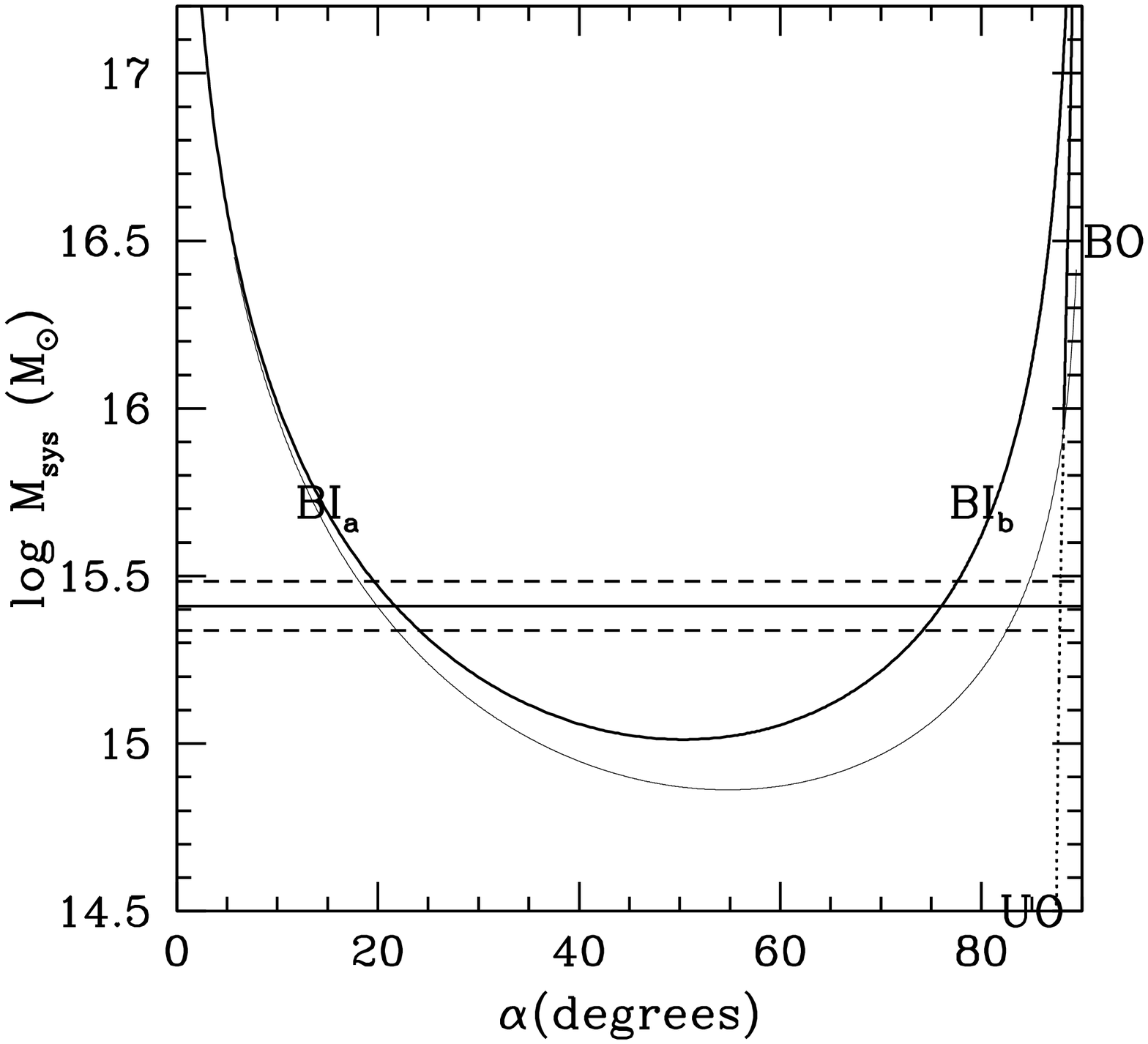}}
\caption
{System mass vs. projection angle for bound and unbound solutions of
the two--body model applied to A115N and A115S subclusters (solid and
dotted curves, respectively, see text).  The thin curve separates the
bound and unbound regions according to the Newtonian criterion (above
and below the curve, respectively).  The horizontal lines give the
observational values of the mass system and its uncertainties.}
\label{figbim}
\end{figure}

First, we consider the Newtonian criterion for gravitational binding
stated in terms of the observables as ${\rm V}_{\rm r}^2D\leq2GM_{\rm
sys}\sin^2\alpha \cos\alpha$, where $\alpha$ is the projection angle
between the plane of the sky and the line connecting the centres of
two clumps.  The faint curve in Fig.~\ref{figbim} separates the bound
and unbound regions according to the Newtonian criterion (above and
below the curve, respectively). Considering the value of $M_{\rm
sys}$, the A115N+S system is bound between $20\degree$ and
$84\degree$; the corresponding probability, computed considering the
solid angles (i.e., $\int^{84^{\rm o}}_{20^{\rm o}}
\cos\,\alpha\,d\alpha$), is 65\%.  We also consider the implemented
criterion ${\rm V}_{\rm r}^2D\leq2GM\sin^2\alpha_{\rm v}
\cos\alpha_{\rm d}$, which introduces different angles $\alpha_{\rm
d}$ and $\alpha_{\rm v}$ for projection of distance and velocity, not
assuming strictly radial motion between the clumps (Hughes et
al. \cite{hug95}).  We obtain a binding probability of 60\%.

Then, we apply the analytical two--body model introduced by Beers et
al. (\cite{bee82}) and Thompson (\cite{tho82}; see also Lubin et
al. \cite{lub98} for a recent application).  This model assumes radial
orbits for the clumps with no shear or net rotation of the
system. Furthermore, the clumps are assumed to start their evolution
at time $t_0=0$ with separation $d_0=0$, and are moving apart or
coming together for the first time in their history; i.e. we are
assuming that we are seeing the cluster prior to merging (at the time
t=11.106 Gyr at the cluster redshift).  The bimodal model solution
gives the total system mass $M_{\rm sys}$ as a function of $\alpha$
(e.g., Gregory \& Thompson \cite{gre84}). Fig.~\ref{figbim} compares
the bimodal--model solutions with the observed mass of the system,
which is the most uncertain observational parameter.  The present
bound outgoing solutions (i.e. expanding), BO, are clearly
inconsistent with the observed mass.  The possible solutions span
these cases: the bound and present incoming solution
(i.e. collapsing), BIa and BIb, and the unbound--outgoing solution, UO.
For the incoming case there are two solutions because of the ambiguity
in the projection angle $\alpha$.  We compute the probabilities
associated to each solution assuming that the region of $M_{\rm sys}$
values between uncertainties are equally probable for individual
solutions: $P_{\rm BIa}\sim83.5\%$, $P_{\rm BIb}\sim16.5\%$, $P_{\rm
UO}\sim6\times 10^{-4}$\%.

Between the two possible bound solutions, $\alpha \sim $ 21 and 76
degrees, the second one is rather unlikely since it means a distance
of $\sim 3.7$ \h between A115N and A115S, i.e. well larger than the
virial radii, while the second solution means a distance of $\sim 1.0$
\h in agreement with a certain degree of interaction.  When assuming
$\alpha =21$ degrees the present colliding velocity is very large
($\sim 4400$ \kss) and the cluster cores will cross after 0.08 Gyr.

The characterization of the dynamics of A115 using these models is
affected by several limitations. For instance, possible underestimates
of the masses -- e.g., if the subclusters extend outside the virial
radius -- lead to binding probabilities larger than those computed
above.  The models do not take the mass distribution in the
subclusters into account when the separation of the subclusters is
comparable with their size (i.e. at small $\alpha$) and do not
consider the possible effect of small, low velocity groups. Moreover,
the two--body model breaks down in a regime where A115N and A115S
subclusters are already strongly interacting.

Finally, the two--body model does not consider the possibility of an
off--axis merger as suggested by the X--ray surface brightness
distribution (Gutierrez \& Krawczynski \cite{gut05}).  For an
analytical treatment which assumes that A115N and A115S are in a
circular Keplerian orbit with the orbital plane perpendicular to the
LOS, we refer to those authors.

As for the low velocity groups, we consider a group corresponding to
WGAP1 likely centred around BCM--D and possibly interacting with the
A115N+S complex. The values of the relevant observable quantities for
the two--clumps system are then: the relative LOS velocity in the rest
frame, $V_{\rm r}=3376$ \ks (as recovered from WGAP1 and the mean
velocity of CORE--A and CORE--B cluster rest frame); the projected
linear distance between the two clumps, $D=0.89$ \h (as recovered from
BCM--D and the mean position between BCM--A and --B); the mass of the
system obtained by adding the mass of a $\sigma_{\rm v}\sim500$ \ks
group to A115N+S, i.e.  log$M_{\rm sys}=15.4427_{-0.0625}^{+0.0625}$.
According to the Newtonian criterion for
gravitational binding we find a binding probability of 38\%.  The
bimodal model indicates that the group is infalling onto the A115N+S
complex with a merging axis intermediate between the LOS and the plane
of sky (i.e., $\alpha\sim $40--60 degrees).

\subsection{Active galaxies}
\label{sec:act}

It has been suggested that cluster--cluster collisions may trigger star
formation in cluster galaxies (Bekki \cite{bek99}; Moss \& Whittle
\cite{mos00}; Girardi \& Biviano \cite{gir02} and references
therein). Caldwell \& Rose (\cite{cal97}) noticed that post--starburst
galaxies are frequently found in clusters with evidence of past
collision events. Here we analyse possible segregations between
passive and active galaxy populations.

Out of 85 cluster members we classify 48 galaxies finding 34
``passive'' galaxies (``k'' in Table~\ref{tab1}) and 14 ``active''
galaxies (i.e., 9 ``k+a''/``a+k'' and 5 ``e'' galaxies, respectively),
see also Fig.~\ref{figcmspettri}. 

The velocity distribution of passive galaxies differs from that of
active galaxies at the 99.47\% c.l.  according to the KS--test,
active galaxies having a larger mean velocity (see
Table~\ref{tabv}). With present data this difference seems due to both
Balmer absorption lines (``k+a''/``a+k'') and emission lines ``e''
galaxies.  This suggests that the galaxy population of the high
velocity subcluster (A115N) is very active with respect to other A115
galaxies.  Indeed, we find that two out three classified galaxies in
CORE--A are active galaxies (in particular BCM--A), while no active
galaxy is found in CORE--B. That A115S has a rich, red galaxy
population is also found by Rakos et al. (\cite{rak00}) suggesting
that this system is evolutionarily ``developed''.  Analysing the 2D
distributions we only find a marginal difference between passive and
Balmer absorption lines galaxies (at the 92\% c.l. according to the
2dKS--test) with three out of nine Balmer absorption lines galaxies
closely located to BCM--A.  Assuming that galaxy activity is connected
with the merging phenomena, our results point out that A115N
subcluster is more affected than A115S in agreement with our finding
that A115N is less massive than A115S (see Sect.~\ref{sec:mass}).

\section{Summary \& conclusions}
\label{sec:sum}

We present the results of the dynamical analysis of the rich, X--ray
luminous, and hot cluster of galaxies A115, showing a binary apparency
(A115N and A115S) and containing a diffuse arc--shape radio emission,
connected to A115N.  This emission is considered a very anomalous
relic since elongated relics are generally located at the cluster
periphery, and with the largest dimension roughly perpendicular to the
cluster radial orientation.

Our analysis is based on new redshift data for 115 galaxies, measured
from spectra obtained at the TNG in a cluster region within a radius
of 15\arcmm$\times$ 20\arcm.  We also use new photometric data
obtained at the INT telescope in a field larger than
30\arcmm$\times$30\arcm.

We select 85 cluster members around $z\sim0.193$ and compute a global
LOS velocity dispersion of galaxies, $\sigma_{\rm v}=1362_{-108}^{+126}$ 
\kss.

Our analysis confirms the presence of two structures of cluster--type
well recognizable in the plane of the sky and shows that they differ
by $\sim 2000$ \ks in LOS velocity.  The northern, high velocity
subcluster (A115N) is likely centred on the second brightest cluster
galaxy (BCM--A, coincident with radio source 3C28) and the northern
X--ray peak.  The southern, low velocity subcluster (A115S) is likely
centred on the first brightest cluster galaxy (BCM--B) and the
southern X--ray peak.  We estimate that A115S is slightly dynamically
more important than A115N having a velocity dispersion of $\sigma_{\rm
v}=900-1100$ \ks vs. $\sigma_{\rm v}=750-850$ \kss.  The virial mass
estimates for the two subclusters are $M_{\rm N}(<{\rm R}_{\rm
vir,N}=2.2-2.5\,\hhh)=8.0$--11.6 \mquaa$\,$and $M_{\rm S}(<{\rm R}_{\rm
vir,S}=2.6-2.9\,\hhh)=13.8$ --18.9 \mquaa.

Moreover, we find evidence for two small groups at low velocities.  
In fact, the galaxy distribution obtained from our INT data shows
the presence of two concentrations (C and D) in addition to A115N and
A115S (A and B).  The C group surrounds the third brightest cluster
member and coincides with the clump already found by Beers et
al. (\cite{bee83}; see also our Dressler--Schectman analysis).  The D
group surrounds another bright galaxy (BCM--D in this paper) and
likely coincides with a faint X--ray emission shown by Chandra data
(Gutierrez \& Krawczynski \cite{gut05}).

Considering the complex structure of A115 we estimate a global cluster
virial mass of 2.2--3.5 \mquii.

Our results agree with a pre--merging scenario where A115N and A115S
are colliding with a LOS impact velocity $\Delta {{\rm v}_{\rm rf}}
\sim 1600$ \ks.  The most likely solution to the two--body problem
suggests that the merging axis lies at $\sim 20$ degrees from the
plane of the sky and that the cores will cross after $\sim 0.1$ Gyr.

In our scenario where 1) A115S is more massive than A115N; 2) the
dynamically important axis is the axis connecting the two merging
subclusters; 3) this axis is perpendicular to the largest dimension of
the relic, the anomaly of the A115 radio diffuse source is likely
overcome supporting its relic nature (Govoni et al. \cite{gov01b}).

\begin{acknowledgements}
We thank Andrea Biviano for useful discussions.\\
This publication is based on observations made on the island of La
Palma with the Italian Telescopio Nazionale Galileo (TNG), operated by
the Fundaci\'on Galileo Galilei -- INAF (Istituto Nazionale di
Astrofisica), and with the Isaac Newton Telescope (INT), operated by
the Isaac Newton Group (ING), in the Spanish Observatorio of the Roque
de Los Muchachos of the Instituto de Astrofisica de Canarias.\\ 
This publication also makes use of data obtained from the Chandra data
archive at the NASA Chandra X--ray centre
(http://cxc.harvard.edu/cda/).\\
This work was partially
supported by a grant from the Istituto Nazionale di Astrofisica (INAF,
grant PRIN--INAF2006 CRA ref number 1.06.09.09).
\end{acknowledgements}

\addtocounter{figure}{-15}
\begin{figure*}
\centering
\includegraphics[width=15.5cm,angle=180]{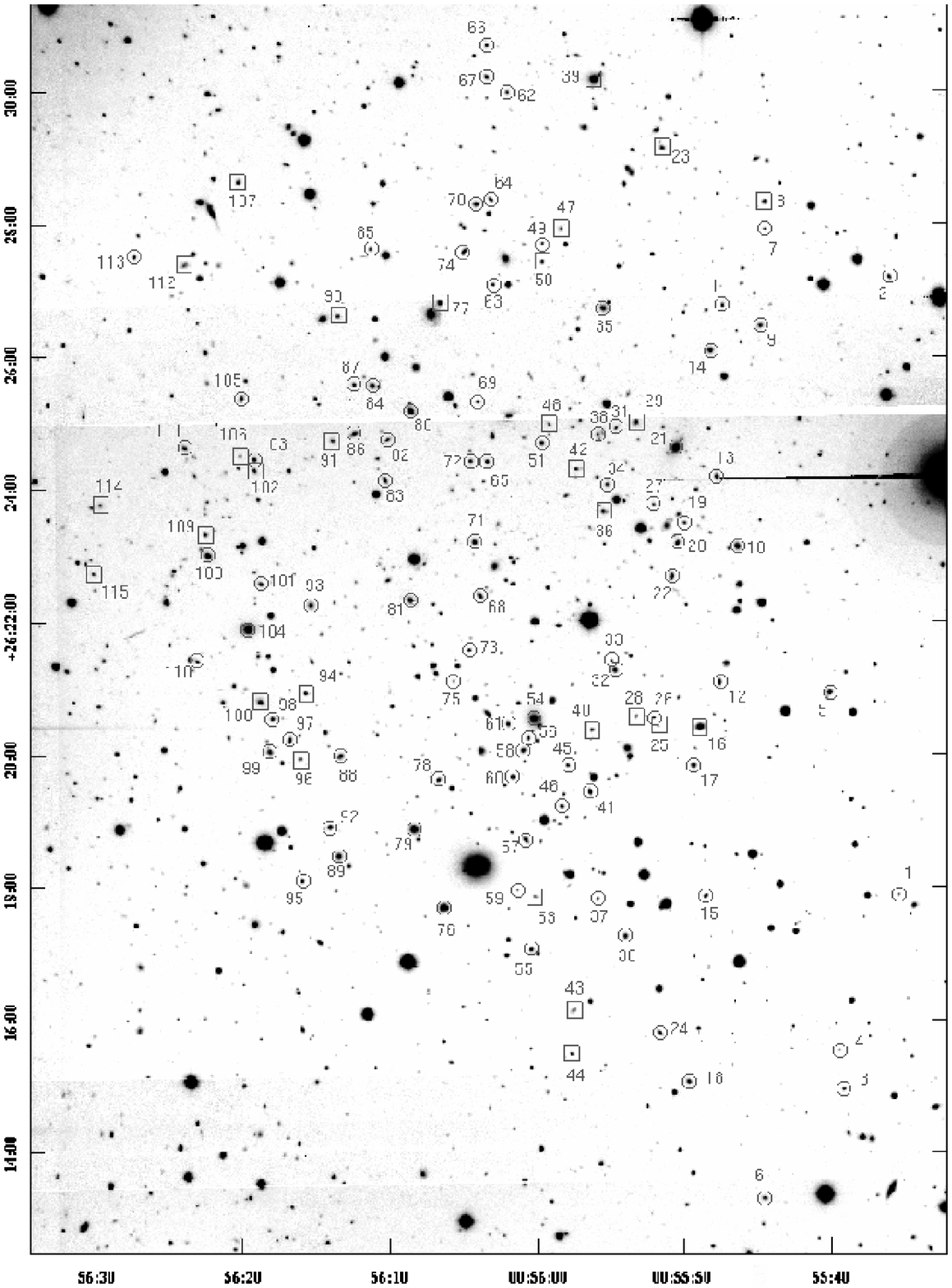}
\caption{$R$--band image of A115 taken with the WFC camera of the
INT. Targets with successful velocity measurements are labeled as in
Table~\ref{tab1}. Circles and boxes indicate cluster members and non
member galaxies, respectively.}
\label{figimage}
\end{figure*}

\begin{figure*}[!ht] \centering
\includegraphics[width=15.5cm,angle=180]{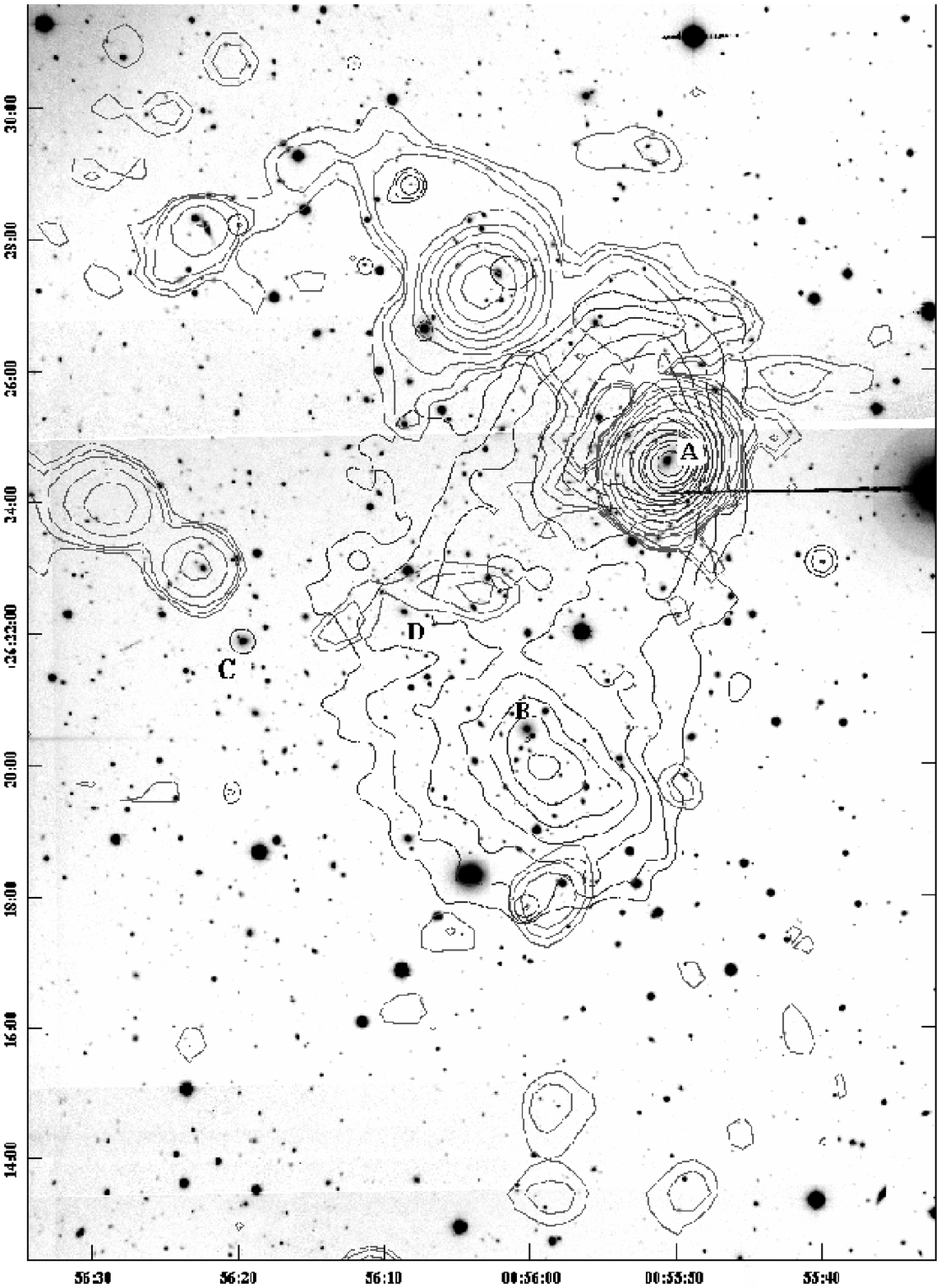}
\caption{$R$--band image of the cluster A115 with, superimposed, the
smoothed contour levels of the Chandra X--ray image ID\#3233 (photons
in the energy range 0.5--5 keV; blue) and the contour levels of the
NVSS (Condon et al. \cite{con98}) radio image (red). Letters A, B, C
and D indicate the three brightest cluster members each one
corresponding to Beers et al. (\cite{bee83}) clumps and a fourth
bright galaxy (see text).}
\label{figisofote}
\end{figure*}

\addtocounter{table}{-2}

\begin{table*}[!ht]
\caption[]{This data will be availabe just after publication in A\&A.}
         \label{catalogue}
              $$ 
           \begin{array}{r c c c c c r c}
            \hline
            \noalign{\smallskip}
            \hline
            \noalign{\smallskip}

\mathrm{ID} & \mathrm{IAU\,\,ID} & \mathrm{\alpha},\mathrm{\delta}\,(\mathrm{J}2000)  & \mathrm{B} & \mathrm{R}  & \mathrm{v} & \mathrm{\Delta}\mathrm{v} & \mathrm{SC} \\
  & & 00^{\mathrm{h}}      , +26^{\mathrm{o}}    &  &  &\,\,\,\,\,\,\,\mathrm{(\,km}&\mathrm{s^{-1}\,)}\,\,\,&  \\
            \hline
            \noalign{\smallskip}   

		 \noalign{\smallskip}			      
	     \hline					     
	     \noalign{\smallskip} 
	     \hline		       
	  \end{array}
     $$ 
\end{table*}
\addtocounter{table}{-1}

\begin{table*}[!ht]
          \caption[ ]{Continued.}
     $$ 
           \begin{array}{r c c c c c r c}
            \hline
            \noalign{\smallskip}
            \hline
            \noalign{\smallskip}

\mathrm{ID} & \mathrm{IAU\,\,ID}& \mathrm{\alpha},\mathrm{\delta}\,(\mathrm{J}2000)  & \mathrm{B} & 
\mathrm{R}  & \mathrm{v} & \mathrm{\Delta}\mathrm{v} & \mathrm{SC} \\
  & & 00^{\mathrm{h}}      , +26^{\mathrm{o}}    &  &  &\,\,\,\,\,\,\,\mathrm{(\,km}&\mathrm{s^{-1}\,)}\,\,\,&  \\
            \hline
            \noalign{\smallskip} 
		\noalign{\smallskip}			     
	    \hline					    
	    \noalign{\smallskip}			    
	    \hline					    
        \end{array}\\
\label{tab1}
     $$ 
\end{table*}



\begin{thebibliography}{}

\bibitem[1989]{abe89} Abell, G. O., Corwin, H. G. Jr., \& Olowin, R. P. 1989, \apjs, 70, 1

\bibitem[1994]{ash94} Ashman, K. M., Bird, C. M. \& Zepf, S. E. 1994, \aj, 108, 2348

\bibitem[1994]{bar94} Bardelli, S., Zucca, E., Vettolani, G., et al. 1994, \mnras, 267, 665

\bibitem[2002]{bar02} Barrena, R., Biviano, A., Ramella, M., Falco, E. E., \& Seitz, S. 2002, \aap, 386, 816

\bibitem[2007]{bar07} 
Barrena, R., Boschin, W., Girardi, M., \& 
Spolaor, M. 2007, \aap, in press (astro-ph/0701833)


\bibitem[1990]{bee90} Beers, T. C., Flynn, K. \& Gebhardt, K. 1990, \aj, 100, 32

\bibitem[1991]{bee91} Beers, T. C., Forman, W., Huchra, J. P., Jones, C., \& Gebhardt, K. 1991, \aj, 102, 1581

\bibitem[1982]{bee82} Beers, T. C., Geller, M. J., \& Huchra, J. P. 1982, \apj, 257, 23

\bibitem[1983]{bee83} Beers, T. C., Geller, M. J., \& Huchra, J. P. 1983, \apj, 264, 365

\bibitem[1999]{bek99} Bekki, K. 1999, \apj, 510, L15

\bibitem[1996]{ber96} Bertin, E. \& Arnouts, S. 1996, \aaps, 117, 393

\bibitem[1993]{bir93} Bird, C. M., \& Beers, T. C. 1993, \aj, 105, 1596

\bibitem[2004]{biv04} Biviano, A., \& Katgert 2004, \aap, 424, 779

\bibitem[2002]{biv02} Biviano, A., Katgert, P., Thomas, T., \& Adami, C. 2002, \aap, 387, 8

\bibitem[2002]{boh02} B\"ohringer, H., \& Schuecker, P. 2002, in
``Merging Processes in Galaxy Clusters'', eds. L. Feretti,
I. M. Gioia, \& G. Giovannini, Kluwer Ac. Pub., The Netherlands:
Observational signatures and statistics of galaxy cluster mergers

\bibitem[2004]{bos04} Boschin, W., Girardi, M., Barrena, R., et al. 2004, \aap, 416, 839

\bibitem[2006]{bos06} Boschin, W., Girardi, M., Spolaor, M., \& Barrena, R. 2006, \aap, 449, 461

\bibitem[2001]{bul01} Bullock, J. S., Kolatt, T. S., Sigad, Y., et al. 2001, \mnras, 321, 559

\bibitem[2002]{buo02} Buote, D. A. 2002, in ``Merging Processes in
Galaxy Clusters'', eds. L. Feretti, I. M. Gioia, \& G. Giovannini,
Kluwer Ac. Pub., The Netherlands: Optical Analysis of Cluster Mergers

\bibitem[1994]{bur94} Burns, J. O., Roettiger, K., Ledlow, M., \& Klypin, A. 1994, \apj, 427, L87

\bibitem[1982]{bur82} Burstein, D., \& Heiles, C. 1982, \aj, 87, 1165

\bibitem[1997]{cal97} Caldwell, N. \& Rose, J. A. 1997, \aj, 113, 492

\bibitem[1997]{car97} Carlberg, R. G., Yee, H. K. C., \& Ellingson, E. 1997, \apj, 478, 462

\bibitem[1998]{con98} Condon, J. J., Cotton, W. D., Greisen, E. W., et al. 1998, \aj, 115, 1693

\bibitem[1976]{cou76} Cousins, A. W. J., 1976, Mem. R. Astr. Soc, 81, 25

\bibitem[1999]{cra99} Crawford, C. S., Allen, S. W., Ebeling, H., Edge, A. C., \& Fabian, A. C. 1999, \mnras, 306, 857

\bibitem[1980]{dan80} Danese, L., De Zotti, C., \& di Tullio, G. 1980, \aap, 82, 322

\bibitem[1996]{den96} den Hartog, R., \& Katgert, P. 1996, \mnras, 279, 349 

\bibitem[2001]{dia01} Diaferio, A., Kauffmann, G., Balogh, M. L., et al. 2001, \mnras, 323, 999

\bibitem[2004]{dol04} Dolag, K., Bartelmann, M., Perrotta, F., et al. 2004, \aap, 416, 853

\bibitem[1988]{dre88} Dressler, A., \& Shectman, S. A. 1988, \aj, 95, 985

\bibitem[1999]{dre99} Dressler, A., Smail, I., Poggianti, B. M., et
al. 1999, \apjs, 122, 51

\bibitem[1994]{ell94} Ellingson, E., \& Yee, H. K. C. 1994, \apjs, 92, 33

\bibitem[2002]{ens02} Ensslin, T. A., \& Br\"uggen, M. 2002, \mnras, 331, 1011

\bibitem[1996]{fad96} Fadda, D., Girardi, M., Giuricin, G., Mardirossian, F., 
\& Mezzetti, M. 1996, \apj, 473, 670

\bibitem[1987]{fas87} Fasano, G., \& Franceschini, A. 1987, \mnras, 225, 155

\bibitem[1999]{fer99} Feretti, L. 1999, MPE Report No. 271

\bibitem[2005]{fer05} Feretti, L. 2005, X--Ray and Radio Connections 
(eds. L. O. Sjouwerman and K. K. Dyer) Published electronically by NRAO,
http://www.aoc.nrao.edu/events/xraydio Held 3--6 February 2004 in Santa
Fe, New Mexico, USA

\bibitem[2002]{fer02a} Feretti, L. 2002, The Universe at Low Radio
Frequencies, Proceedings of IAU Symposium 199, held 30 Nov -- 4 Dec
1999, Pune, India.  Edited by A. Pramesh Rao, G. Swarup, and
Gopal--Krishna, 2002., p.133

\bibitem[1984]{fer84} Feretti, L., Gioia, I. M., Giovannini, G., et al. 1984, \aap, 139, 50

\bibitem[1981]{for81} Forman, W., Bechtold, J., Blair, W., et al. 1981, \apjl,243, 133

\bibitem[2004]{fuj04} Fujita, Y., Sarazin, C. L., Reiprich, T. H., et al. 2004, \apj, 616, 157

\bibitem[2002]{gio02} Giovannini, G., \& Feretti, L. 2002, in
``Merging Processes in Galaxy Clusters'', eds. L. Feretti,
I. M. Gioia, \& G. Giovannini, Kluwer Ac. Pub., The Netherlands:
Diffuse Radio Sources and Cluster Mergers

\bibitem[1987]{gio87} Giovannini, G., Feretti, L., \& Gregorini,
L. 1987, \aaps, 69, 171

\bibitem[1993]{gio93} Giovannini, G., Feretti, L., Venturi T., Kim, K.--T., \& Kronberg, P. P. 1993, \apj, 406, 399 

\bibitem[1999]{gio99} Giovannini, G., Tordi, M., \& Feretti, L. 1999, New Astronomy, 4, 141

\bibitem[2002]{gir02} Girardi, M., \& Biviano, A. 2002, in ``Merging
Processes in Galaxy Clusters'', eds. L. Feretti, I. M. Gioia, \&
G. Giovannini, Kluwer Ac. Pub., The Netherlands: Optical
Analysis of Cluster Mergers

\bibitem[1993]{gir93} Girardi, M., Biviano, A., Giuricin, G., Mardirossian, F., \& Mezzetti, M. 1993, \apj, 404, 38

\bibitem[2006]{gir06} Girardi, M., Boschin, W., \& Barrena, R. 2006, \aap, 455, 45

\bibitem[2005]{gir05} Girardi, M., Demarco, R., Rosati, P., \& Borgani, S. 2005, \aap, 442, 29

\bibitem[1996]{gir96} Girardi, M., Fadda, D., Giuricin, G. et al. 1996, \apj, 457, 61

\bibitem[1998]{gir98} Girardi, M., Giuricin, G., Mardirossian, F., Mezzetti, M., \& Boschin, W. 1998, \apj, 505, 74

\bibitem[2001]{gir01} Girardi, M., \& Mezzetti, M. 2001, \apj, 548, 79

\bibitem[2001a]{gov01a} Govoni, F., Ensslin, T. A., Feretti, L., \& Giovannini, G. 2001a, \aap, 369, 441

\bibitem[2001b]{gov01b} Govoni, F., Feretti, L., Giovannini, G., et al. 2001b, \aap, 376, 803

\bibitem[1984]{gre84} Gregory, S. A., \& Thompson, L. A. 1984, ApJ, 286, 422

\bibitem[1992]{gul92} Gullixson, C. A. 1992, in ``Astronomical CCD Observing and Reduction Techniques'' (ed. S. B. Howell), ASP Conf. Ser., 23, 130 

\bibitem[2005]{gut05} Gutierrez, K., \& Krawczynski, H. 2005, \apj, 619,161

\bibitem[1982]{han82} Hanisch, R. J. 1982, \aap, 116, 137

\bibitem[1995]{hug95} Hughes, J. P., Birkinshaw, M., \& Huchra, J. P. 1995, \apjl, 448, 93

\bibitem[1992]{mal92} Malumuth, E. M., Kriss, G. A., Dixon, W. Van Dyke, Ferguson, H. C., \& Ritchie, C. 1992, \aj, 104, 495

\bibitem[1953]{joh53} Johnson, H. L., \&  Morgan, W. W. 1953, \apj, 117, 313

\bibitem[1999]{jon99} Jones, C., \& Forman, W. 1999, \apj, 511, 65

\bibitem[1992]{ken92} Kennicutt, R. C. 1992, \apjs, 79, 225

\bibitem[1960]{lim60} Limber, D. N., \& Mathews, W. G. 1960, \apj, 132, 286

\bibitem[2004]{lop04} L\'opez--Cruz, O.,  Barkhouse, W. A., \& Yee, H. K. C. 2004, \apj, 614, 679

\bibitem[1998]{lub98} Lubin, L. M., Postman, M., \& Oke, J. B. 1998, \aj, 116, 643

\bibitem[2001]{mar01} Markevitch, M., \& Vikhlinin, A. 2001, \apj, 563, 95 

\bibitem[1996]{men96} Menci, N., \& Fusco--Femiano, R. 1996, \apj, 472, 46

\bibitem[2000]{mos00} Moss, C., \& Whittle, M. 2000, \mnras, 317, 667

\bibitem[1986]{nag86} NAG Fortran Workstation Handbook, 1986 (Downers
Grove, IL: Numerical Algorithms Group)

\bibitem[1997]{nav97} Navarro, J. F., Frenk, C. S., \& White, S. D. M. 1997, \apj, 490, 493

\bibitem[1996]{pin96} Pinkney, J., Roettiger, K., Burns, J. O., \& Bird, C. M. 1996, \apjs, 104, 1

\bibitem[1993]{pis93} Pisani, A. 1993, \mnras, 265, 706

\bibitem[1996]{pis96} Pisani, A. 1996, \mnras, 278, 697

\bibitem[1997]{pog97} Poggianti, B. M. 1997, \aaps, 122, 399

\bibitem[1999]{pog99} Poggianti, B. M., Smail, I., Dressler, A., et al. 1999, \apj, 518, 576

\bibitem[1992]{pre92} Press, W. H., Teukolsky, S. A., Vetterling,
W. T., \& Flannery, B. P. 1992, in Numerical Recipes (Second Edition),
(Cambridge University Press)

\bibitem[2000]{rak00} Rakos, K. D., Schombert, J. M., Odell, A. P., \& Steindling, S. 2000, \apj, 540, 715

\bibitem[1975]{ril75} Riley, J. M., \& Pooley, G. G. 1975, \mnras, 80, 105

\bibitem[1997]{roe97} Roettiger, K., Loken, C., \& Burns, J. O. 1997, \apjs, 109, 307

\bibitem[1994]{rot94} R\"ottgering, H., Snellen, I., Miley, G., et al. 1994, \apj, 436, 654

\bibitem[2002]{sar02} Sarazin, C. L. 2002, in ``Merging Processes in Galaxy 
Clusters'', eds. L. Feretti, I. M. Gioia, \& G. Giovannini, Kluwer Ac. Pub., 
The Netherlands: The Physics of Cluster Mergers

\bibitem[2005]{sau05} Sauvageot, J. L., Belsole, E., \& Pratt, G. W. 2005, \aap, 444, 673

\bibitem[1987]{sch87} Schombert, J. M. 1987, \apjs, 64, 643

\bibitem[2001]{sch01} Schuecker, P., B\"ohringer, H., Reiprich, T. H., \& Feretti, L. 2001, \aap, 378, 408

\bibitem[1965]{sha65} Shapiro, S. S., \& Wilk, M. B. 1965, Biometrika, 52, 591

\bibitem[1999]{shi99} Shibata, R., Honda, H., Ishida, M., Ohashi, T., \& Yamashita, K. 1999, \apj, 524, 603

\bibitem[1986]{the86} The, L. S., \& White, S. D. M. 1986, \aj, 92, 1248

\bibitem[1982]{tho82} Thompson, L. A. 1982, in IAU Symposium 104,   
Early Evolution of the Universe and the Present Structure,
ed. G.O. Abell and G. Chincarini, Dordrecht:Reidel

\bibitem[1979]{ton79} Tonry, J., \& Davis, M. 1979, \aj, 84, 1511

\bibitem[1993]{tri93} Tribble, P. 1993, \mnras, 263, 31

\bibitem[1978]{wai78} Wainer, H., \&  Schacht, S. 1978, Psychometrika, 43, 203

\bibitem[1997]{whi97} White, D. A., Jones. C., \& Forman, W.  1997, \mnras, 292, 419

\bibitem[2005]{yua05} Yuan, Q. R., Yan, P. F., Yang, Y. B., \& Zhou, X. 2005, Chinese Journal of Astronomy and Astrophysics, 5, 126

\bibitem[1990]{zab90} Zabludoff, A. I., Huchra, J. P., \& Geller, M. J. 1990, \apjs, 74, 1

\end{thebibliography}
\end{document}